\documentclass[a4paper,12pt]{article}

\usepackage{latexsym,amssymb,fancyhdr,lastpage}
\usepackage{graphicx,amsfonts}
\usepackage{times,mathptmx,bm,amsmath}
\usepackage{dcolumn}

\usepackage{graphicx,float,placeins}
\usepackage{mathrsfs}
\usepackage{tikz}
\usepackage{soul,color}
\usepackage[normalem]{ulem}

\usetikzlibrary{decorations.markings}
\usetikzlibrary{decorations.pathmorphing}
\usetikzlibrary{fit}
\usetikzlibrary{calc,arrows,patterns}
\usepgflibrary{plotmarks}

\usepackage{paralist}

\newcommand{\mb}{\mathbf}

\newcommand{\tr}{\mathrm{tr}\,}

\newcommand{\be}{\begin{equation}}
\newcommand{\ee}{\end{equation}}

\newcommand{\im}{I_{\rm m}}

\newcommand{\sL}{\mathscr{L}}

\newcommand{\al}{\alpha}

\newcommand{\e}{{\rm e}}

\renewcommand{\theequation}{\thesection.\arabic{equation}}
\numberwithin{equation}{section}

\title{On the complex singularities of the inverse Langevin function}
\author{S. R. Rickaby\thanks{Email: stephen.r.rickaby@gmail.com}, 
N. H. Scott\thanks{Email: n.scott@uea.ac.uk}\\
School of Mathematics, University of East Anglia,\\
 Norwich Research Park, Norwich NR4 7TJ
 }

\date{}

\begin{document}
\maketitle

\thispagestyle{fancy} \lhead{\emph{IMA Journal of Applied Mathematics}   (2018) {\bf 83},  1092--1116.    \\   
doi: 10.1093/imamat/hxy046\\ 
\emph{Advance access 17 September 2018}\\
arXiv: 
}
\chead{13 May 2020}
\rhead{Page\  \thepage\ of\ \pageref{LastPage}}
\cfoot{}

\emph{[Received on 1 March 2018; revised on 22 August 2018; accepted on 24 August 2018]}

\begin{abstract}
We study the inverse Langevin function $\mathscr{L}^{-1}(x)$ because of its importance in modelling limited-stretch elasticity where the stress and strain energy become infinite as a certain maximum strain is approached, modelled here by $x\to1$.  The only real singularities of the inverse Langevin function $\mathscr{L}^{-1}(x)$ are two simple poles at $x=\pm1$ and we see how to remove their effects either multiplicatively or additively.   In addition, we find that $\mathscr{L}^{-1}(x)$ has an infinity of complex singularities.   Examination of the Taylor series about the origin of $\mathscr{L}^{-1}(x)$ shows that the four complex singularities nearest the origin are equidistant from the origin and  have the same strength; we develop a new algorithm for finding these four complex singularities.  Graphical illustration seems to point to these complex singularities being of a square root nature.  An exact analysis then proves these are square root branch points.

\vspace{0.0mm}\noindent
\end{abstract}

\section{Introduction} 
\label{sec:1}

In modelling the stress softening of rubber or rubber-like materials many authors in the past have utilized  either the James \& Guth \cite{james} three-chain model, the Wang \& Guth \cite{wang} four-chain model, the Arruda \& Boyce \cite{arruda} eight-chain model or the Wu \& Van der Giessen \cite{wu} full network model.  Z\'{u}\~{n}iga \& Beatty \cite{zuniga2002} and Beatty \cite{beatty2003} give descriptions of these models, all of which are  closely related to the original Kuhn \& Gr\"un \cite{kuhn} single-chain model.  The chains referred to here are arbitrarily orientated chains of connected molecules.  When these polymer  chains are stretched to their maximum extent there results a maximum stretch of the rubber sample in the direction of extension.    The stress response and strain energy then become infinite as this maximum stretch is reached.  Such a model of rubber elasticity is said to be a limited-stretch  elastic model. 

The Langevin function is defined by
\be\label{1.1x}
x=\mathscr{L}(y)=\coth y - 1/y
\ee
which has a removable singularity at $y=0$ and is defined on the domain $-\infty<y<\infty$.
All the above models, however, involve the inverse Langevin function defined by
\be\label{1.2x}  
y=\mathscr{L}^{-1}(x) 
\ee
defined on the domain $-1<x<1$ with range $-\infty<y<\infty$.
These are both odd functions  though on physical grounds $y$ and $x$ may be restricted to be positive.  It is the property $\mathscr{L}^{-1}(x)\to\infty$ as $x\to 1$ that makes the inverse Langevin function so useful in modelling limited-stretch elasticity.

\pagestyle{fancy}
\lhead{Complex singularities of $\mathscr{L}^{-1}(x)$} \chead{Page\  \thepage\ of\ \pageref{LastPage}} \rhead{S. R. Rickaby, N. H. Scott}
\lfoot{} \cfoot{} \rfoot{}

The Arruda \& Boyce \cite{arruda} eight-chain model has proved to be the most successful of these models, both theoretically and experimentally.  However, Beatty \cite{beatty2003} has shown that quite remarkably the Arruda \& Boyce \cite{arruda} stress response holds in general in isotropic nonlinear elasticity for a full network model of arbitrarily orientated molecular chains.  Thus the eight-chain cell structure is unnecessary and so the inverse Langevin function has an important role to play in  the theory of the  finite isotropic elasticity of rubber and rubber-like materials.

The inverse Langevin function cannot be expressed in closed form and so many approximations to it have been devised and applied to many models of rubber elasticity.  Perhaps the simplest is to consider the Taylor series but this does not converge over the whole domain of definition $-1<x<1$ of the inverse Langevin function, see \cite{itskov} and \cite{itskov2011}.   Cohen \cite{cohen} derived an approximation based upon a $[3/2]$ Pad\'e approximant of the inverse Langevin function which has been widely used and is fairly accurate over the whole domain of definition.  Many more accurate approximations have been devised, see for example,  \cite{darabi, jedynak2015, jedynak2017, kroger, marchi, nguessong, puso, rickaby5, treloar1975}.  However, rather than finding further approximations to the inverse Langevin function, we emphasise in this paper how to find the approximate positions  of its singularities in the complex plane.  We are also able to find exactly the positions and nature of these singularities.

This paper is structured as follows.  In Section \ref{sec:2} we give a brief account of non-linear isotropic elasticity theory as applied to the limited-stretch theory of elasticity of, for example, \cite{arruda} and \cite{beatty2003}.   In Section \ref{sec:3} we define and discuss the Langevin and inverse Langevin functions, giving Taylor series for both, and identify the real singularities of the inverse Langevin function to be two simple poles. We see that the effects of these poles may be removed either  multiplicatively or additively.  In Section \ref{sec:4} we give approximate methods for analysing the Taylor series of functions with four complex singularities equidistant from the origin.  First, in Section \ref{sec:4.1}  we extend  the methods of Mercer \& Roberts \cite{mercer} and Hunter \& Guerrieri \cite{hunter} in the two-singularity case to the present situation of four complex singularities equidistant from the origin and develop an algorithm for estimating the four complex  singularities with the smallest radius of convergence. Then, in Section \ref{sec:4.2} a continued fraction method is used to calculate the poles and zeros of the Taylor series approximation and also the method of Pad\'{e} approximants is considered.
In Section \ref{sec:5.1} there is a graphical representation of the four branch cut singularities nearest the origin found using the methods developed in subsection \ref{sec:4.1}.   In Section \ref{sec:5.2} these methods are used to  discuss and illustrate the branch cut singularities which are the next-nearest to the origin and in Section \ref{sec:5.3} there is a brief discussion of Euler's method for removing the four nearest singularities to infinity.  An exact analysis of the complex singularities of the inverse Langevin function is given in Section \ref{sec:6}.  The complex singularities are identified as square root branch points and the first 100 are given in Tables \ref{table:2} and \ref{table:3} correct to 15 significant figures.
Finally, a discussion of the results is given in Section \ref{sec:7}.

\section{Non-linear isotropic incompressible elasticity for rubber-like materials and polymers} 
\label{sec:2}

The Cauchy stress in an incompressible isotropic elastic material is given by
\be \label{2.1x}
\mb{T} = -p\mb{I}+\beta \mb{B} +\beta_{-1}\mb{B}^{-1}
\ee
where $p$ is an arbitrary pressure and $\mb{B}=\mb{F}\mb{F}^{\rm T}$ is the left Cauchy-Green strain tensor with $\mb{F}$ denoting the deformation gradient.  The response functions are given in terms of the strain energy $W$ by
\be \label{2.2x} \beta  = 2\frac{\partial W}{\partial I_1},\quad \beta_{-1}  = -2\frac{\partial W}{\partial I_2} \ee
where 
\be \label{2.3x}
 I_1=\tr\,\mb{B} = \lambda_1^2+\lambda_2^2+\lambda_3^2, \quad
 I_2=\tr\,\mb{B}^{-1} = \lambda_1^{-2}+\lambda_2^{-2}+\lambda_3^{-2}
 \ee
are the first two principal invariants of  $\mb{B}$ given in terms of the principal stretches $\{\lambda_1, \lambda_2, \lambda_3 \}$.  Because of incompressibility the third principal invariant is given by $I_3=\det\mb{B}= \lambda_1^2\lambda_2^2\lambda_3^2=1$.   We are assuming no dependence on  $I_2$, in common with all the models discussed in the previous section,  and so must  take $\beta_{-1}=0$ and $\beta=\beta(I_1)$.  Therefore, throughout this paper, the Cauchy stress (\ref{2.1x}) reduces to
\be\label{2.4x} \mb{T} = -p\mb{I}+\beta \mb{B}, \ee
where the stress response $\beta$ is given by Eq.\@ (\ref{2.2x})$_1$.

Beatty \cite{beatty2008} describes two approaches for modelling limited-stretch elasticity.  The first approach limits the greatest of the three principal stretches by imposing a maximum stretch $\lambda_{\rm m}$ which occurs when the polymer chains are fully extended. The second approach limits the value of the first principal invariant $I_1$ to a maximum value denoted by $\im$ which similarly occurs when the polymer chains are fully extended. From the experimental observations of Dickie \& Smith \cite{dickie} and the theoretical results discussed by  Beatty \cite{beatty2008},    we may conclude that limiting polymer chain extensibility is governed by $\im$ alone so that $\lambda_{\rm m}$ need not be mentioned.
   Therefore, $I_1$ is restricted by
\be\label{2.5x}  3\leq I_1\leq\im. \ee

For future convenience we introduce the new variable
\be\label{2.6x} x=\sqrtsign{\frac{I_1}{I_m}},\quad\mbox{restricted by\quad}
x_0\leq x< 1, \ee
where $x_0=\sqrtsign{{3}/{I_m}}$ is the value of $x$ in the undeformed state, where $I_1=3$.

In terms of the inverse Langevin function  the Arruda-Boyce stress  response function is
\be\label{2.7x}   \beta = \mu\,\mathscr{L}^{-1}(x)/3x,   \ee
see Arruda \& Boyce \cite{arruda} or Beatty \cite{beatty2003}, where $\mu$ is a shear modulus.  As remarked before, \cite{beatty2003} has shown the general applicability of the response function (\ref{2.7x}) in  full network isotropic elasticity.   This stress response depends on only two material constants, the shear modulus $\mu$ and the maximum value  $\im$ of the first principal invariant $I_1$. 

We can integrate $\beta$ given by (\ref{2.7x}) in order to find the strain energy
\[ W  
  = \frac{ \mu\im}{3}\int\mathscr{L}^{-1}(x)\,dx  
  = \frac{ \mu\im}{3} \left( x\mathscr{L}^{-1}(x)+\log \left( \frac{\mathscr{L}^{-1}(x)}{\sinh \mathscr{L}^{-1}(x)} \right)\right) - h_0   \]
where $h_0$ is a constant chosen so that $W=0$ when $x=x_0$.

Both  stress and strain energy become infinite as $x\to 1$, i.e., as $I_1\to \im$, as expected in limited-stretch elasticity.

\section{Properties of the Langevin and inverse Langevin functions} 
\label{sec:3}

The Langevin function defined at (\ref{1.1x}) has Taylor series 
\begin{equation}
\mathscr{L}(y)=\frac{1}{3}y-\frac{1}{45}y^3+\frac{2}{945}y^5-\frac{1}{4725}y^7+\frac{2}{93555}y^9-\frac{1382}{638512875}y^{11}+\cdots 
\label{3.1x}
\end{equation}
and the inverse Langevin function has Taylor series
\begin{equation}
\mathscr{L}^{-1}(x)=3x+\frac{9}{5}x^3+\frac{297}{175}x^5+\frac{1539}{875}x^7+\frac{126117}{67375}x^9+\frac{43733439}{21896875}x^{11}
+\cdots\, .
\label{3.2x}
\end{equation}
Itskov et al. \cite{itskov2011} describe an efficient method for calculating the Taylor series for an inverse function and use it  to calculate the inverse Langevin function to 500 terms, the first 59 being presented in their paper.  Itskov et al. \cite{itskov2011} also estimated the radius of convergence of this series to be $r_1\approx 0.904$.

It can be shown that the only real singularities of $ \mathscr{L}^{-1}(x) $ are two simple poles,  at $x=\pm 1$, each with residue $-1$. 

\subsection{Multiplicative removal of the simple poles of $ \mathscr{L}^{-1}(x) $} 
\label{sec:3.1}

We can remove these two simple poles   by considering instead the reduced inverse Langevin function $f(x)$ of \cite{rickaby5} defined by
\begin{equation}\label{3.3x}
\begin{aligned}
f(x) & =\frac{1-x^2}{3x}\mathscr{L}^{-1}(x) \\
 &= 
1-\frac{2}{5}x^2-\frac{6}{175}x^4+\frac{18}{875}x^6+\frac{2538}{67375}x^8+\frac{915138}{21896875}x^{10}+\cdots,
\end{aligned}
\end{equation}
which may be termed a multiplicative removal of the poles of  $ \mathscr{L}^{-1}(x) $. 
In fact, $f(x)$ remains finite at $x=\pm1$ as can be seen by using (\ref{1.1x}) to write $x$ in (\ref{3.3x})$_1$ in terms of $y$ and replacing the limit $x\to 1$ by the equivalent  limit $y\to\infty$
to show that $f(\pm 1)= \frac23$ and further that $f'(\pm 1)= \mp \frac13$, see Rickaby \& Scott \cite{rickaby5} for more details.  This is illustrated in Figure \ref{fig:1w}$(a)$.

Rickaby \& Scott \cite[Equation (57)]{rickaby5}  took the first two terms of the series (\ref{3.3x})$_2$ to obtain the approximation
\be \label{3.4x} \mathscr{L}^{-1}(x) \approx \frac{3x}{1-x^2}(1-\tfrac25x^2)  \ee
to the inverse Langevin function and employed it in their model of cyclic stress softening of an orthotropic material in pure shear, see Rickaby \& Scott \cite{rickaby6}. Kr\"oger \cite[Equation (F.5)]{kroger} misquotes (\ref{3.4x}) and so deduces wrongly that this model does not have the correct oddness in $x$. 

\subsection{Additive removal of the simple poles of $ \mathscr{L}^{-1}(x) $} 
\label{sec:3.2}

The simple poles of $ \mathscr{L}^{-1}(x) $   give a total pole contribution of
\[ \frac{-1}{x+1} + \frac{-1}{x-1} = \frac{2x}{1-x^2} \]
so that we can decompose $ \mathscr{L}^{-1}(x) $ additively as
\[
 \mathscr{L}^{-1}(x) = \frac{2x}{1-x^2} +g(x)
\]
where we define
\be
g(x)= - \frac{2x}{1-x^2} +  \mathscr{L}^{-1}(x).
\label{3.5x}
\ee
 Now
$ - 2x/(1-x^2)  = -2(x+x^3+x^5+\cdots) $
and so from  (\ref{3.2x}) and  (\ref{3.5x}) we obtain
\be
g(x) = x-\frac15x^3-\frac{53}{175}x^5 -\frac{211}{875}x^7 - \frac{8633}{67375}x^9 
- \frac{60311}{21896875}x^{11} +\cdots
\label{3.6x}
\ee
where each coefficient in (\ref{3.6x}) is exactly 2 less than the corresponding coefficient in (\ref{3.2x}). 
It is clear from (\ref{3.6x}) that $g(0)=0$ and $g'(0)=1$.  By using (\ref{1.1x}) to write $g(x)$ in terms of $y$, as with $f(x)$ above, and taking the limit $y\to\infty$ we find that $g(1)=\frac12$ and $g'(1)= - \frac14$.  Thus $g(x)$ remains finite at $x=\pm1$. 

\begin{figure}[h] 
\centerline{
\begin{tikzpicture}
\node (0,0) {\includegraphics[scale=0.9]{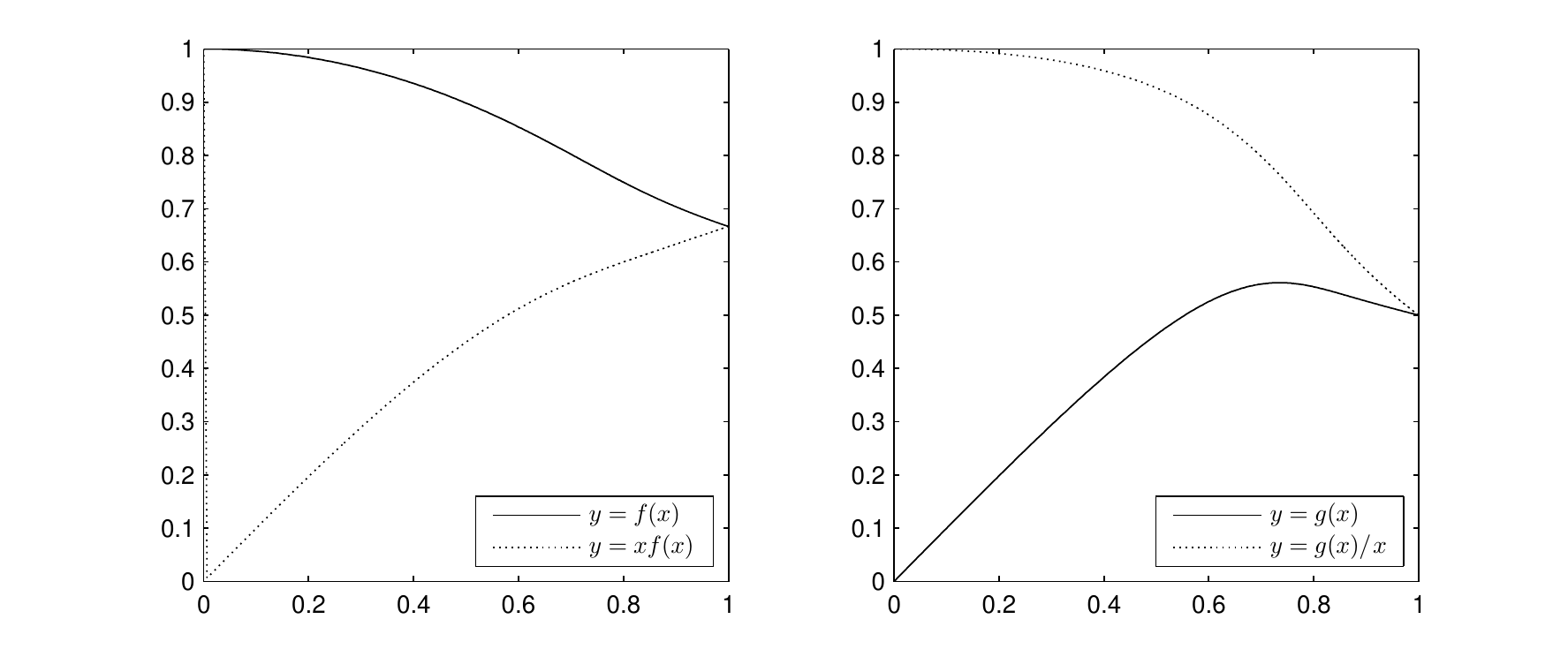}};
\draw  (-6.8,0.1) node [rotate=90] {\fontsize{10}{10} $y$};
\draw  (0.40,0.1) node [rotate=90] {\fontsize{10}{10} $y$};
\draw  (-3.3,-3.3) node {\fontsize{10}{10} $x$};
\draw  (-3.3,-3.8) node {\fontsize{10}{10} $(a)$};
\draw  (3.9,-3.3) node {\fontsize{10}{10} $x$};
\draw  (3.9,-3.8) node {\fontsize{10}{10} $(b)\;\;$};
\end{tikzpicture}}  \vspace{-5pt}
\caption{$(a)$ The functions $f(x)$ and $xf(x)$.  $(b)$ The functions $g(x)$ and $h(x) = g(x)/x$.}
\label{fig:1w}
\end{figure}
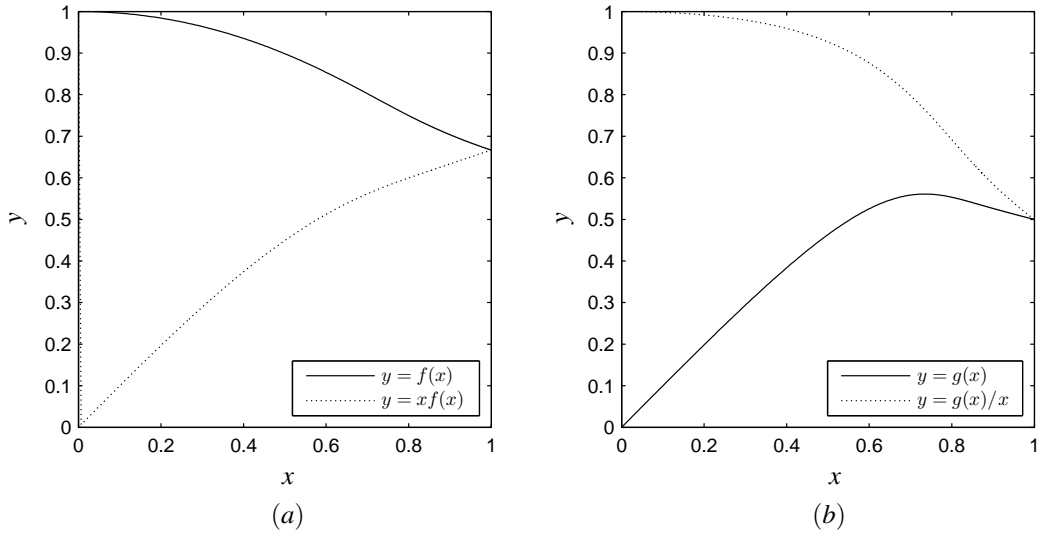

We define a new function $h(x)$ by
\be \label{3.7x}
\begin{aligned}
h(x) = \frac{g(x)}{x} &= - \frac{2}{1-x^2} +  \frac{\mathscr{L}^{-1}(x)}{x}, \\
  & = 1-\frac15x^2-\frac{53}{175}x^4 -\frac{211}{875}x^6 - \frac{8633}{67375}x^8 
- \frac{60311}{21896875}x^{10} +\cdots
\end{aligned}
\ee
an  even function in $x$  satisfying $h(0) = 1$, $h'(0) = 0$, $h(1) = \frac12$ and $h'(1) = -\frac34$.  The functions $g(x)$ and $h(x)$ are illustrated in Figure \ref{fig:1w}$(b)$.

The Taylor series for $f(x)$, $g(x)$ and $h(x)$ each have the same radius of convergence as that for $\mathscr{L}^{-1}(x)$.  The Taylor series for $h(x)$ is given in the Appendix as far as the term in~$x^{448}$.

We may take the first term of (\ref{3.6x}) to obtain the approximation 
\be\label{3.8x}
 \mathscr{L}^{-1}(x) \approx \frac{2x}{1-x^2}+x
 \ee
 which has the correct singular behaviour as $x\to\pm1$ and the correct value of  $3x$ as $x\to 0$.  This model is, in fact, exactly the same as Cohen \cite{cohen}, as observed by Rickaby \& Scott  \cite{rickaby5}.

\section{Approximate analysis for four complex singularities equidistant from the origin} 
\label{sec:4}

The signs of the coefficients in the series expansions for  $\mathscr{L}^{-1}(x)$, $f(x)$,  $g(x)$ and $h(x)$ each settle down to repeating patterns of length 17, either $9+$ followed by $8-$ signs, or vice versa.  The repeating pattern for $\mathscr{L}^{-1}(x)$ begins at the term $x^{75}$, for $f(x)$  at  the term $x^{34}$, for $g(x)$ at the term  $x^{25}$, and for $h(x)$ at the term  $x^{24}$, indicating that the pole contributions to $\mathscr{L}^{-1}(x)$ have a noticeable  effect on the convergence of its series.  

Each of these series is real and so any singularities not on the real line must occur in complex conjugate pairs.
If the pattern of signs consists of a cycle of length $N$ with $M$ changes of sign then from \cite[p. 145]{Hinch} the pair of singularities have arguments $\pm360M/N$ degrees.  If we consider, for example, the even function $h(x)$ defined at (\ref{3.7x}) as a series in $x^2$ then the cycle has $N=17$ and $M=1$ but the arguments $\pm360M/N$ degrees must be halved to give the arguments in $x$.  Therefore we expect the argument of the singularity of $h(x)$  nearest the origin in the first quadrant to be
\be \label{4.1x}  \theta_1 = 180/17\approx 10.59^\circ.  \ee

\subsection{Extension of the methods of Hunter \& Guerrieri \cite{hunter} and Mercer \& Roberts \cite{mercer}} 
\label{sec:4.1}

We extend the methods of Hunter \& Guerrieri \cite{hunter} and Mercer \& Roberts \cite{mercer} for a single pair of complex conjugate singularities to the present situation where there are four singularities of equal strength equidistant from the origin.

We consider the complex function $h(z)$ defined by (\ref{3.7x}) which  is even in $z$. 
If $z_1=r_1\e^{i\theta_1}$ in the first quadrant is a singular point of $h$ then because the series for $h$ is has only real coefficients the complex conjugate $\bar{z}_1=r_1\e^{-i\theta_1}$ must also be a singular point.  Because of evenness $-z_1$ and $-\bar{z}_1$ are also singular points.   Thus we have the four singular points
\be \label{4.2x}
z=\pm r_1 \e^{\pm i\theta_1}, 
\ee
with $\theta_1$ given by (\ref{4.1x}), all at distance $|z_1|=r_1$ from the origin.  We have yet to determine~$r_1$. 

Mercer \& Roberts \cite[(A.1)]{mercer} show how to model a function with a pair of complex conjugate singularities.  We extend this idea to the case of the four singularities, at  $z=\pm r \e^{\pm i\theta}$, in order to model the even function $h(z)$ defined by (\ref{3.7x}):
\be \label{4.3x}
h(z) = 
\frac{1}{4}  \left(1 - \frac{z}{r \e^{ i\theta} } \right)^{\alpha} 
+ \frac{1}{4}  \left(1 - \frac{z}{r \e^{ - i\theta} } \right)^{\alpha} 
+  \frac{1}{4} \left(1 + \frac{z}{r \e^{ i\theta} } \right)^{\alpha} 
+ \frac{1}{4}  \left(1 + \frac{z}{r \e^{ - i\theta} } \right)^{\alpha} ,
\ee
so that $h(0)= 1$, as expected.  
Using the binomial expansion we obtain
\begin{align} \label{4.4x}
\left(1 - \frac{z}{r \e^{ i\theta} } \right)^{\alpha}  &=
\sum_{n=0}^\infty  \binom{\alpha}{n}  (-1)^n r^{-n}e^{ - in\theta}z^n 
= \sum_{n=0}^\infty  \frac{\Gamma(n-\alpha)}{n!\Gamma(-\alpha)} r^{-n}e^{ - in\theta}z^n 
\end{align}
where $\Gamma(n)=(n-1)!$ is the gamma function.  

By replacing $\theta$ by $-\theta$ in (\ref{4.4x}) and adding the two series we obtain
\[
\left(1 - \frac{z}{r \e^{ i\theta} } \right)^{\alpha} + \left(1 - \frac{z}{r \e^{ - i\theta} } \right)^{\alpha} 
= 2 \sum_{n=0}^\infty  \frac{\Gamma(n-\alpha)}{n!\Gamma(-\alpha)} r^{-n}\cos(n\theta)z^n.
\]
We may obtain the last two terms of (\ref{4.3x}) by replacing $\theta$ by $\theta + \pi$ in (\ref{4.4x}), so that $\cos(n\theta)$ is replaced by $(-1)^n\cos(n\theta)$ and the odd powers of $z$ in (\ref{4.3x}) cancel out leaving only even powers.  Then (\ref{4.3x}) becomes
\be \label{4.5x}
h(z) =  \sum_{n=0, 2, 4, \cdots}^\infty  \frac{\Gamma(n-\alpha)}{n!\Gamma(-\alpha)} r^{-n}\cos(n\theta)z^n
= \sum_{m=0}^\infty  a_{2m}z^{2m},
\ee
where 
\be \label{4.6x}
a_{2m} = \frac{\Gamma(2m-\alpha)}{(2m)!\Gamma(-\alpha)} r^{-2m}\cos(2m\theta).
\ee

By making use of the identity
\[
\cos(2m\theta) - 2\cos 2\theta \cos(2m\theta-2\theta) + \cos(2m\theta-4\theta)=0, \quad m=1, 2, 3, \dots
\]
we can show that any three consecutive coefficients (\ref{4.6x})  of the series (\ref{4.5x}) satisfy exactly  the equation
\begin{align} \label{4.7x}
r^4a_{2m} -  2&\cos 2\theta \frac{(2m-1-\alpha)(2m-2-\alpha)}{2m(2m-1)}r^2a_{2m-2} \nonumber
 \\[2mm]  
&+ \frac{(2m-1-\alpha)(2m-2-\alpha)(2m-3-\alpha)(2m-4-\alpha)}{2m(2m-1)(2m-2)(2m-3)}a_{2m-4}=0
\end{align}
for $m=2,3, \dots$.  
For $2m$ large it might suffice to approximate (\ref{4.7x}) by 
\be \label{4.8x}
r^4a_{2m} -  2\cos 2\theta \left( 1 - \frac{2+2\alpha}{2m}\right) r^2a_{2m-2} 
+ \left( 1 - \frac{4+4\alpha}{2m}\right) a_{2m-4}=0
\ee
for $m=2,3, \dots$, which agrees with (\ref{4.7x}) as far as terms $O(1/2m)$.  

For known approximate values of the coefficients $a_{2m}$  equation (\ref{4.7x}) or (\ref{4.8x}) can be used as a basis for approximating the position of the singularity $z=r\e^{i\theta}$ and its index~$\alpha$.  We shall consider the function $h(x)$ defined by (\ref{3.7x}) with its Taylor series (\ref{A1}) furnishing the coefficients $a_{2m}$.  We may regard either (\ref{4.7x}) or (\ref{4.8x}) as an equation for the three unknowns $r$, $\cos2\theta$ and $\alpha$ for each value of $m$.  Taking (\ref{4.7x}) for three consecutive values of $m$ gives a system of three  equations in the three unknowns which can be solved simultaneously for  $r$, $\cos2\theta$ and $\alpha$.  For example, the line $2m=262$ of Table \ref{table:1} was obtained by solving  equations (\ref{4.7x}) for $2m=258, 260, 262$, and so on.  From Table \ref{table:1} we see that $r$  is converging   to the value  $r_1\approx 0.905$, close to the value $r_1\approx 0.904$ of \cite{itskov2011}   and $\theta$ is converging to the value $\theta_1$ given by (\ref{4.1x}).  Thus $z=r\e^{i\theta}$ is converging to $z_1=r_1\e^{i\theta_1}$ given by (\ref{4.2x}).  However, the convergence of $\alpha$ is poor.  Table \ref{table:1a} is constructed in the same way as Table \ref{table:1} except that solutions of the approximate equations   (\ref{4.8x}) are employed instead of solutions of the more accurate equations   (\ref{4.7x}).

\begin{table}[h]  
\centering
\scalebox{0.85}{
\begin{tabular}{cc c c c c c c c c c}
\hline\hline 
$2m$ &  $a_{2m}$  & $a_{2m-2}$ & $a_{2m-4}$  &$r$  &    $\cos 2\theta$ & $\alpha$ &  eq. (\ref{4.7x})  \\  [0.2ex] 
\hline
262&    -1.29624$e$+08&    -2.11799$e$+06&   \phantom{-} 8.55592$e$+07&    0.90424&    0.93240&   \phantom{-} 0.60479&   \phantom{-} 1.19209$e$--07\\
264&    -2.88951$e$+08&    -1.29624$e$+08&    -2.11799$e$+06&    0.90787&    0.93249&    -0.43364&    -6.05360$e$--09\\
266&    -4.61842$e$+08&    -2.88951$e$+08&    -1.29624$e$+08&    0.90502&    0.93242&   \phantom{-} 0.38075&    \phantom{-}0.00000$e$+00\\
268&    -6.18836$e$+08&    -4.61842$e$+08&    -2.88951$e$+08&    0.90483&    0.93242&   \phantom{-} 0.43476&    -6.55651$e$--07\\
270&    -7.20187$e$+08&    -6.18836$e$+08&    -4.61842$e$+08&    0.90475&    0.93242&   \phantom{-} 0.45857&    \phantom{-}0.00000$e$+00\\
272&    -7.19324$e$+08&    -7.20187$e$+08&    -6.18836$e$+08&    0.90469&    0.93242&   \phantom{-} 0.47528&   \phantom{-} 0.00000$e$+00\\
274&    -5.69319$e$+08&    -7.19324$e$+08&    -7.20187$e$+08&    0.90464&    0.93242&   \phantom{-} 0.49116&   \phantom{-} 0.00000$e$+00\\
276&    -2.32412$e$+08&    -5.69319$e$+08&    -7.19324$e$+08&    0.90457&    0.93242&   \phantom{-} 0.51145&   \phantom{-} 0.00000$e$+00\\
278&   \phantom{-} 3.07952$e$+08&    -2.32412$e$+08&    -5.69319$e$+08&    0.90445&    0.93241&   \phantom{-} 0.55042&    \phantom{-}0.00000$e$+00\\
280&   \phantom{-} 1.03392$e$+09&    \phantom{-}3.07952$e$+08&    -2.32412$e$+08&    0.90374&    0.93240&   \phantom{-} 0.76526&    \phantom{-}1.69873$e$--06\\
282&   \phantom{-} 1.88100$e$+09&   \phantom{-} 1.03392$e$+09&    \phantom{-}3.07952$e$+08&    0.90524&    0.93243&   \phantom{-} 0.30835&    -8.94070$e$--07\\
284&   \phantom{-} 2.72986$e$+09&   \phantom{-} 1.88100$e$+09&    \phantom{-}1.03392$e$+09&    0.90487&    0.93243&   \phantom{-} 0.42057&    -3.93391$e$--06\\
286&   \phantom{-} 3.40563$e$+09&   \phantom{-} 2.72986$e$+09&    \phantom{-}1.88100$e$+09&    0.90477&    0.93242&   \phantom{-} 0.45167&   \phantom{-} 0.00000$e$+00\\
288&   \phantom{-} 3.68836$e$+09&   \phantom{-} 3.40563$e$+09&    \phantom{-}2.72986$e$+09&    0.90471&    0.93242&   \phantom{-} 0.46962&    -4.76837$e$--06\\
290&   \phantom{-} 3.33768$e$+09&    \phantom{-}3.68836$e$+09&    \phantom{-}3.40563$e$+09&    0.90466&    0.93242&   \phantom{-} 0.48440&   \phantom{-} 1.23978$e$--05\\
292&   \phantom{-} 2.13308$e$+09&    \phantom{-}3.33768$e$+09&    \phantom{-}3.68836$e$+09&    0.90461&    0.93242&    \phantom{-}0.50062&    -1.43051$e$--05\\
294&    -7.12750$e$+07&    \phantom{-}2.13308$e$+09&    \phantom{-}3.33768$e$+09&    0.90454&    0.93242&   \phantom{-} 0.52538&    \phantom{-}9.53674$e$--06\\
296&    -3.28170$e$+09&    -7.12750$e$+07&    \phantom{-}2.13308$e$+09&    0.90433&    0.93242&   \phantom{-} 0.59257&    \phantom{-}2.62260$e$--06\\
298&    -7.29899$e$+09&    -3.28170$e$+09&    -7.12750$e$+07&    0.90728&    0.93248&    -0.36369&    -8.10623$e$--06\\
300&    -1.16644$e$+10&    -7.29899$e$+09&    -3.28170$e$+09&    0.90494&    0.93243&    \phantom{-}0.39451&   \phantom{-} 3.67165$e$--05\\
\hline 
\end{tabular}}
\caption{ Estimates for the radius of convergence $r$,  the argument $\theta$ and the index $\alpha$ of the first singularity  of  $h(z)$. The final column gives the value of the left hand side of eq. (\ref{4.7x}) when the values from the Table are substituted into it.  } 
\label{table:1} 
\end{table}

\begin{table}[h]  
\centering
\scalebox{0.85}{
\begin{tabular}{cc c c c c c c c c c}
\hline\hline 
$2m$ &  $a_{2m}$  & $a_{2m-2}$ & $a_{2m-4}$  &$r$  &    $\cos 2\theta$ & $\alpha$ &  eq. (\ref{4.8x})\\[0.2ex]  
\hline 
262&   -1.29624$e$+08&   -2.11799$e$+06& \phantom{-}  8.55592$e$+07&  0.90403&   0.93236&   \phantom{-}0.65828&   \phantom{-}0.00000$e$+00\\
264&   -2.88951$e$+08&   -1.29624$e$+08&   -2.11799$e$+06&   0.91694& 	  0.93253&   -3.06739&   -1.52737$e$--07\\
266&   -4.61842$e$+08&   -2.88951$e$+08&   -1.29624$e$+08&   0.90511&  		 0.93241&   \phantom{-}0.35354&   \phantom{-}5.06639$e$--07\\
268&   -6.18836$e$+08&   -4.61842$e$+08&   -2.88951$e$+08&   0.90489&  	 0.93240&  \phantom{-} 0.41620&   -7.74860$e$--07\\
270&   -7.20187$e$+08&   -6.18836$e$+08&   -4.61842$e$+08&   0.90478&  0.93240&   \phantom{-}0.44551&   \phantom{-}0.00000$e$+00\\
272&   -7.19324$e$+08&   -7.20187$e$+08&   -6.18836$e$+08&   0.90471&  0.93240&  \phantom{-} 0.46678&   \phantom{-}0.00000$e$+00\\
274&   -5.69319$e$+08&   -7.19324$e$+08&   -7.20187$e$+08&   0.90464&   0.93240&  \phantom{-} 0.48760&  \phantom{-} 0.00000$e$+00\\
276&   -2.32412$e$+08&   -5.69319$e$+08&   -7.19324$e$+08&   0.90455&  	 0.93239&   \phantom{-}0.51509&  \phantom{-} 0.00000$e$+00\\
278&   \phantom{-}3.07952$e$+08&   -2.32412$e$+08&   -5.69319$e$+08&   	0.90436&  0.93239&  \phantom{-} 0.57114&  \phantom{-} 0.00000$e$+00\\
280&  \phantom{-} 1.03392$e$+09&  \phantom{-} 3.07952$e$+08&   -2.32412$e$+08&   0.90284&   0.93233&   \phantom{-}1.02720&   \phantom{-}0.00000$e$+00\\
282&   \phantom{-}1.88100$e$+09&  \phantom{-} 1.03392$e$+09&  \phantom{-} 3.07952$e$+08&   0.90534&   0.93243&  \phantom{-} 0.27627&   -1.31130$e$--06\\
284&  \phantom{-} 2.72986$e$+09&  \phantom{-} 1.88100$e$+09&  \phantom{-} 1.03392$e$+09&   0.90493&   0.93241&  \phantom{-} 0.39919&   -3.09944$e$--06\\
286&  \phantom{-} 3.40563$e$+09&  \phantom{-} 2.72986$e$+09&   \phantom{-}1.88100$e$+09&   0.90481&   0.93241&  \phantom{-} 0.43683&   \phantom{-}0.00000$e$+00\\
288&   \phantom{-}3.68836$e$+09&  \phantom{-} 3.40563$e$+09&  \phantom{-} 2.72986$e$+09&   0.90473&   0.93241&   \phantom{-}0.45944&   \phantom{-}0.00000$e$+00\\
290&  \phantom{-} 3.33768$e$+09& \phantom{-}  3.68836$e$+09& \phantom{-}  3.40563$e$+09&   0.90467&   0.93241&  \phantom{-} 0.47858&   \phantom{-}1.38283$e$--05\\
292&  \phantom{-} 2.13308$e$+09&  \phantom{-} 3.33768$e$+09&  \phantom{-} 3.68836$e$+09&   0.90460&   0.93241&  \phantom{-} 0.50018&   -9.05991$e$--06\\
294&   -7.12750$e$+07&  \phantom{-} 2.13308$e$+09&   3.33768$e$+09&   0.90450&   0.93240&   \phantom{-}0.53447&   \phantom{-}1.04904$e$--05\\
296&   -3.28170$e$+09&   -7.12750$e$+07&   \phantom{-}2.13308$e$+09&   0.90417&   0.93239&  \phantom{-} 0.63724&   \phantom{-}3.33786$e$--06\\
298&   -7.29899$e$+09&   -3.28170$e$+09&   -7.12750$e$+07&   0.91628&   0.93251&   -3.31514&   -2.89083$e$--06\\
300&   -1.16644$e$+10&   -7.29899$e$+09&   -3.28170$e$+09&   0.90501&   0.93242&   \phantom{-}0.36901&   \phantom{-}3.48091$e$--05\\
\hline 
\end{tabular}}
\caption{ Estimates for the radius of convergence $r$, the argument $\theta$ and the index $\alpha$ of the first singularity  of  $h(z)$.   The final column gives the value of the left hand side of eq. (\ref{4.8x}) when the values from the Table are substituted into it.  } 
\label{table:1a} 
\end{table}

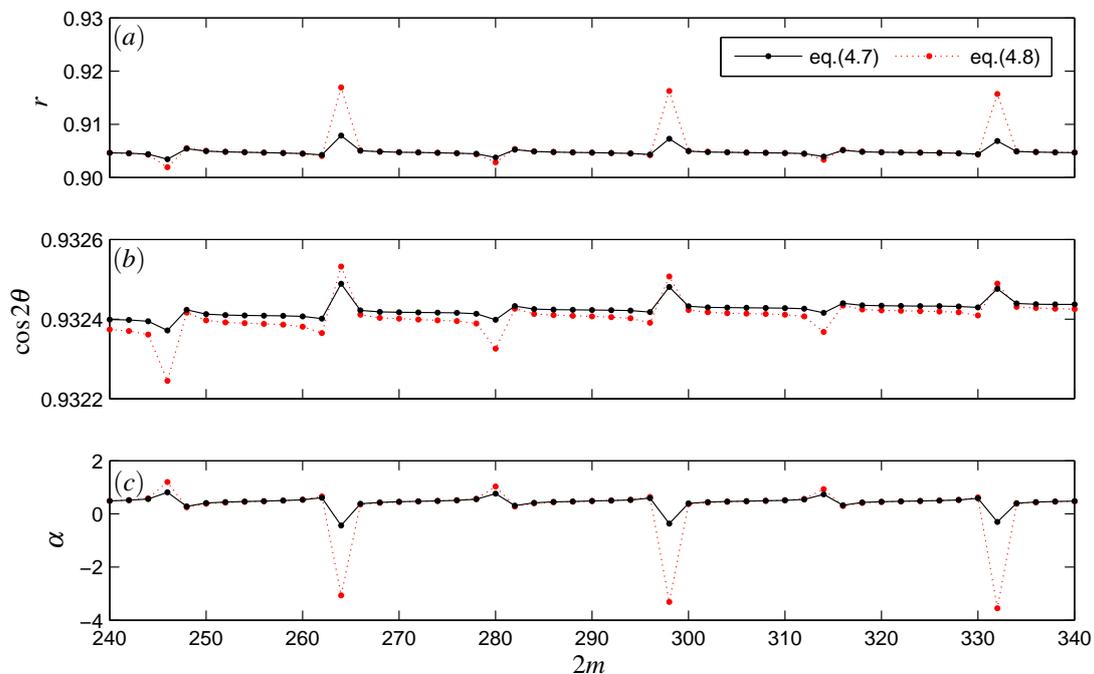
\begin{figure}[H] 
\centerline{
\begin{tikzpicture}
\node (0,0) {\includegraphics[scale=0.845]{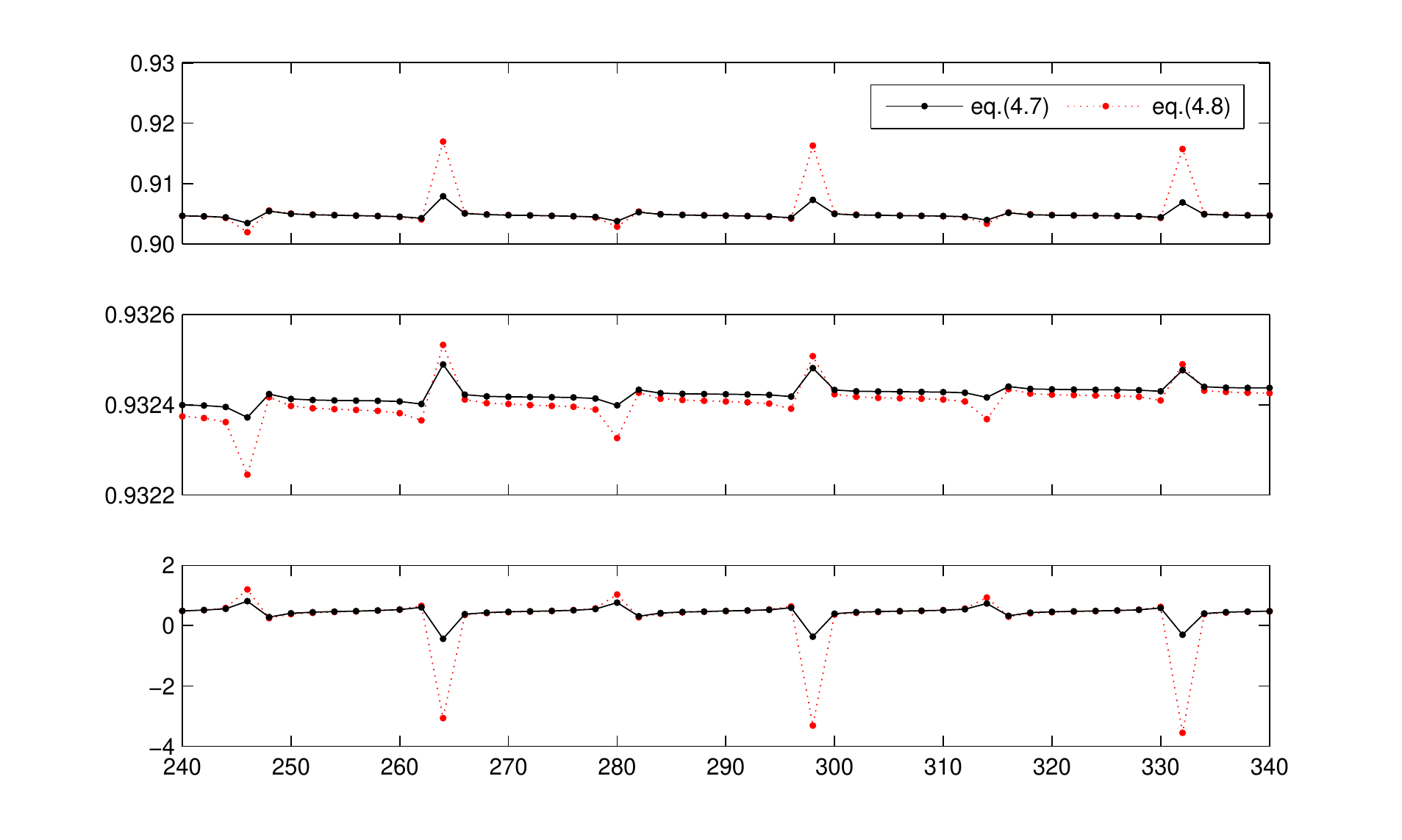}};
\draw  (-6.95,3.0) node [rotate=90] {\fontsize{10}{10} $r$};
\draw  (-7.23,0.1) node [rotate=90] {\fontsize{10}{10} $\cos2\theta$};
\draw  (-6.75,-2.8) node [rotate=90] {\fontsize{10}{10} $\alpha$};
\draw  (0.2,-4.4) node {\fontsize{10}{10} $2m$};
\draw  (-5.85,3.9) node {\fontsize{10}{10} $(a)$};
\draw  (-5.85,0.95) node {\fontsize{10}{10} $(b)$};
\draw  (-5.85,-2.0) node {\fontsize{10}{10} $(c)$};
\end{tikzpicture}}  \vspace{-5pt}
\caption{Plot of the estimates from Tables \ref{table:1} and \ref{table:1a} for ($a$) the radius of convergence $r$, ($b$) the quantity $\cos2\theta$ and ($c$) the index $\alpha$ of the first singularity of $h(z)$  for $2m=240..340$. The cycle of length 17 is apparent.  }
\label{fig:2new} 
\end{figure}

In order to investigate further the convergence exhibited in Tables \ref{table:1} and \ref{table:1a} we plot in Figure \ref{fig:2new} the values of $r$, $\cos2\theta$ and $\alpha$ obtained from eq. (\ref{4.7x}) and (\ref{4.8x}) for $2m=240..340$.  In each subplot we see that there is a cycle of length 17 as predicted at the start of this section.  If the two outliers of each cycle (i.e. for the sequence $2m=262..294$ the outliers would be at $2m=264,280$)  in each subplot are ignored the convergence is seen to be quite good.  In subplots \ref{fig:2new}($a$) and \ref{fig:2new}($b$) the outliers differ from the other values only by a small percentage but by a large amount in subplot \ref{fig:2new}($c$).  It is apparent from subplot \ref{fig:2new}($c$) and Table \ref{table:1a} that the outlier for $\alpha$ for the approximate equation (\ref{4.8x}) is far greater than for the more exact equation (\ref{4.7x}).    If we ignore the outliers occurring at $2m=246,264,280,298,314,332$, then the average percentage error between eq. (\ref{4.7x}) and (\ref{4.8x}) over the range $2m=240..340$ for $r$ is  0.0004\%, for $\cos2\theta$ it is 0.002\%  with the error in $\alpha$ being larger at 1.6\%.
Kr\"oger \cite[Figure B.4.]{kroger} and Jedynak \cite[Figure 2]{jedynak2017} both plot $\log\vert a_{2m}\vert$ and obtain approximately straight lines indicating an exponential increase in coefficient values.  The cycle of length 17 is apparent as is the large outlier in each cycle.  This large outlier is what leads to the large outlier for $\alpha$ as observed in Figure \ref{fig:2new}($c$).

When $m$ is large we can use Stirling's formula to show that 
$ \Gamma(2m-\alpha)/(2m)! \sim (2m)^{-(1+\alpha)}$
so that for large $m$ equation  (\ref{4.6x}) may be approximated by
\be \label{4.9x}
a_{2m} \sim \frac{1}{\Gamma(-\alpha)} (2m)^{-(1+\alpha)} r^{-2m}\cos(2m\theta).
\ee

For large $m$ values close together we can replace (\ref{4.9x}) by 
\be \label{4.10x}
a_{2m} \sim Cr^{-2m}\cos(2m\theta)
\ee
where $C$ can be regarded as constant.  Making use of the identity
\[ \cos^2(2m\theta) - \cos(2m\theta + 2\theta)  \cos(2m\theta - 2\theta)  = \sin^2 2\theta,
\quad m=1,2,3, \ldots , \]
we can use (\ref{4.10x}) to show that
\begin{align*}
 a_{2m}^2 & -  a_{2m+2}a_{2m-2}  
  =  C^2r^{-4m} \sin^2 2\theta, 
 \end{align*}
the right hand side being independent of the rapidly varying term $\cos (2m\theta)$.  Therefore we can  define a quantity $B_{2m}$ by
\be \label{4.11x}
 B_{2m}\equiv \left( \frac{ a_{2m}^2 -  a_{2m+2}a_{2m-2}}{ a_{2m-2}^2  
 -  a_{2m}a_{2m-4}} \right)^{1/4}   =  \frac{1}{r}.
\ee

If in (\ref{4.11x}) we replace the coefficients $a_{2m}$ defined at (\ref{4.6x}) with those given in the Appendix for $h(x)$ defined by (\ref{3.7x}) we obtain from (\ref{4.11x}) the estimate
\be \label{4.12x}
r_{2m}^{\phantom{-1}}  = B_{2m}^{-1}
\ee
 for the radius of convergence $r$.  This follows the method of Mercer \& Roberts \cite[Appendix (A.5)]{mercer} in the two singularity case.

We now need a method of estimating $\theta$ and we follow Mercer \& Roberts \cite[Appendix (A.6)]{mercer} in the two singularity case.  Reverting to  $a_{2m}$ defined at (\ref{4.6x}), we define
\begin{align*}
C_{2m} & = \frac12\left[ \frac{a_{2m-2}\cdot B_{2m}^{2}}{a_{2m}} + 
\frac{a_{2m+2}}{a_{2m}\cdot B_{2m}^{2}} \right]  \\
   & = \frac12\left[ \frac{Cr^{-2m+2}\cos(2m\theta-2\theta) \cdot r^{-2}}{Cr^{-2m}\cos(2m\theta)} 
   + \frac{Cr^{-2m-2}\cos(2m\theta+2\theta)}{Cr^{-2m}\cos(2m\theta) \cdot r^{-2}} \right] \\
   & = \frac12\left[  \frac{\cos(2m\theta-2\theta) + \cos(2m\theta+2\theta)}{\cos(2m\theta)}  \right] \\
   & = \cos 2\theta
\end{align*}
and again the rapidly varying term $\cos(2m\theta)$ has been removed. 
Thus for $h(x)$ defined by (\ref{3.7x}) with Taylor series  (\ref{A1}) we can estimate $\cos2\theta$ to be 
\be\label{4.13x}
\cos2\theta = \lim_{m \to\infty} C_{2m}.
\ee

\begin{figure}[H] 
\centerline{
\includegraphics[scale=0.90]{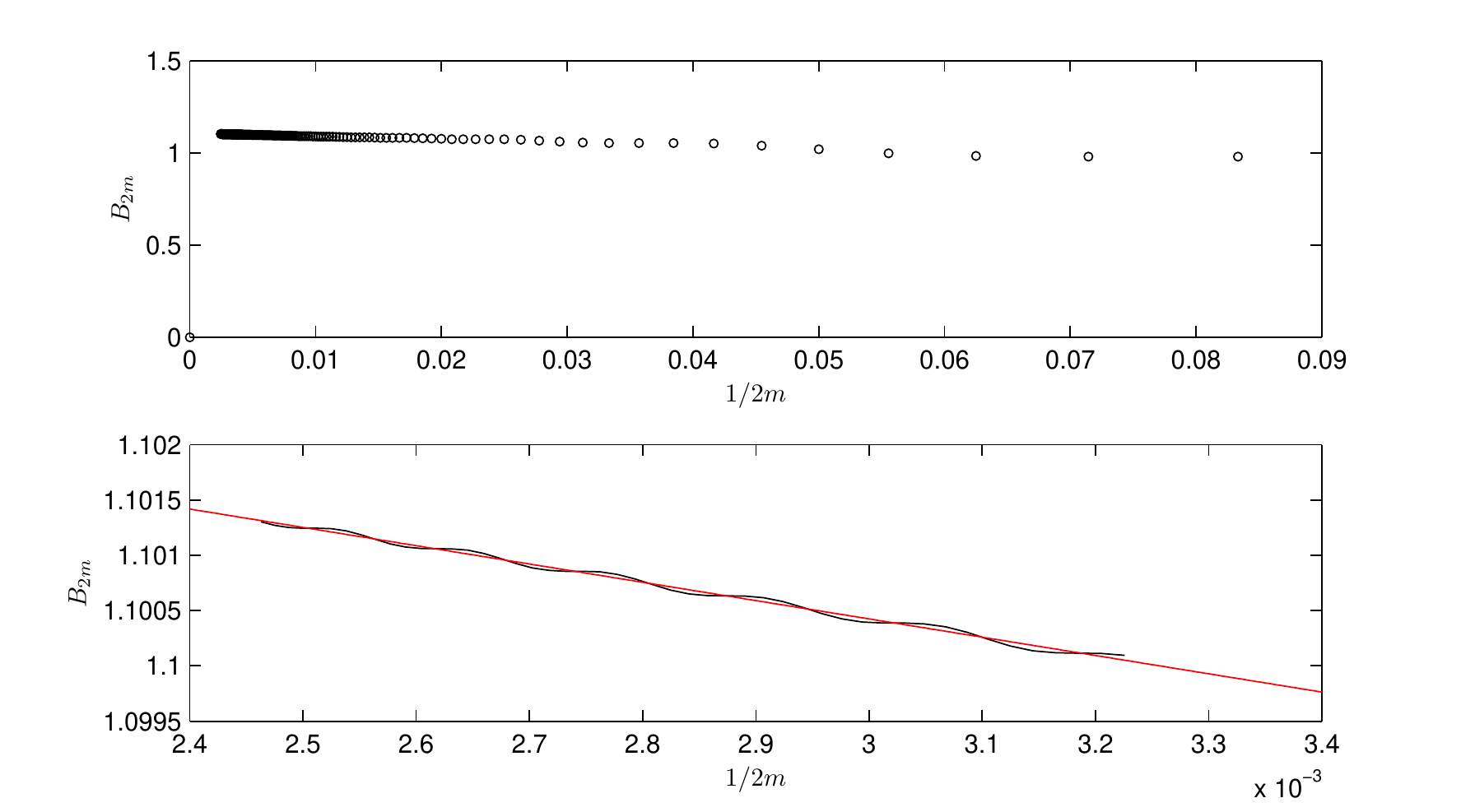}}
\caption{Domb-Sykes plot of Taylor series (\ref{A1}) using equation (\ref{4.14x}).  The lower plot gives an amplified version of the large $m$ regime. }
\label{fig:2w} 
\end{figure}

We now have good methods of estimating $r$ and $\theta$ for the singularity closest to the origin.  However, the order of the singularity $\alpha$ does not appear in the approximation (\ref{4.10x}) and so we shall have to go back to either  the exact expression  (\ref{4.6x}) or the asymptotic result  (\ref{4.9x}).  Arguing from either of these results, keeping terms $O(1/m)$ but discarding terms $O(1/m^2)$, we find that the approximation  (\ref{4.11x}) is replaced by
\be \label{4.14x}
B_{2m} = \frac{1}{r}\left( 1 - \frac{1+\alpha}{2m}  \right).
\ee
In Figure \ref{fig:2w} we plot $B_{2m}$ against $1/2m$ to obtain a standard Domb-Sykes \cite{domb} plot which, as predicted by (\ref{4.14x}), is approximately a straight line for large values of $m$.  The best straight line fit to the data has intercept 1.1053876 and   slope $-1.654448$  giving the estimates
\be \label{4.15x}
r_1\approx 0.9047,\quad   \alpha \approx 0.4967,
\ee
the first agreeing quite well with both the value  $r_1\approx 0.905$ predicted by 
Table \ref{table:1}  and the value $r_1\approx 0.904$ of  Itskov et al. \cite{itskov2011}.  We shall see that (\ref{4.15x})$_2$ gives $\alpha$ correct to 2 dp.

The four singularities (\ref{4.2x}) may thus be approximated by
\be \label{4.16x}
z \approx \pm 0.889 \pm 0.166i.
\ee

\subsection{Continued fraction representations and Pad\'e approximants} 
\label{sec:4.2}

A continued fraction representation can be calculated for any Taylor series and we are able to calculate the poles and zeros of this continued fraction.  These poles and zeros are approximately equal to the poles and zeros calculated using the  Pad\'e approximant to the same Taylor series, see Hinch \cite[pp 151--154] {Hinch} for a discussion of Pad\'e approximants and continued fraction representations. Using Maple the continued fraction representation method takes less computational time than the Pad\'e approximant method and so can be used to calculate a Taylor series expansion of a higher order.   The largest Taylor series that we were able to work with was one of 150 terms.

\begin{figure}[H] 
\centerline{
\includegraphics[scale=0.8]{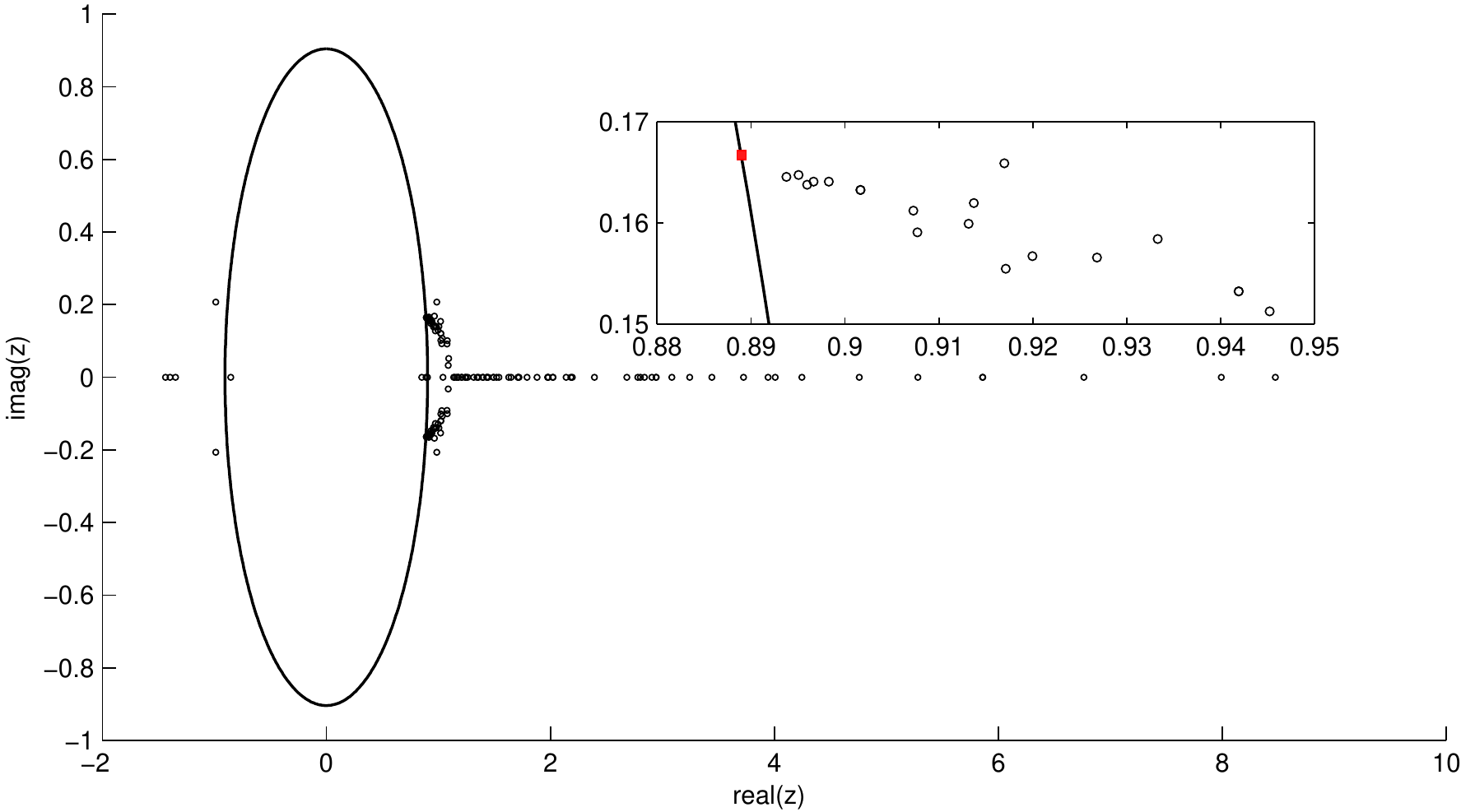}}
\caption{Plot of the singularities found using the continued fraction method, the black circle (which appears as an ellipse because of different axis scalings)  has a radius convergence of $0.905$. The subplot identifies the location of the singularity at $z_1$ in the first quadrant.}
\label{fig:3w} 
\end{figure}

Performing increasing truncations of $5,10,15 \ldots 150$ terms of the continued fraction representation of the Taylor series (\ref{3.3x}) we obtain Figure \ref{fig:3w} in which the black circle (which appears as an ellipse because of different axis scalings) has a radius of convergence of $0.905$.  We have removed  all the spurious pole-zero pairs (Froissart doublets) using the fitting criterion of Gonnet et al.  \cite{trefethen}. 
From Figure \ref{fig:3w} it can be seen that the sequence of poles tend to the circle of convergence of radius $0.905$.   The singularity at $z_1$ in the first quadrant is identified in the subplot in Figure \ref{fig:3w} by the red square. The line of poles radiating out along the real axis at $y=0$ demonstrates the existence of a branch cut at $x \approx 1$, see Hinch \cite[p. 152]{Hinch}.

\section{Graphical representation of the inverse Langevin function} 
\label{sec:5}
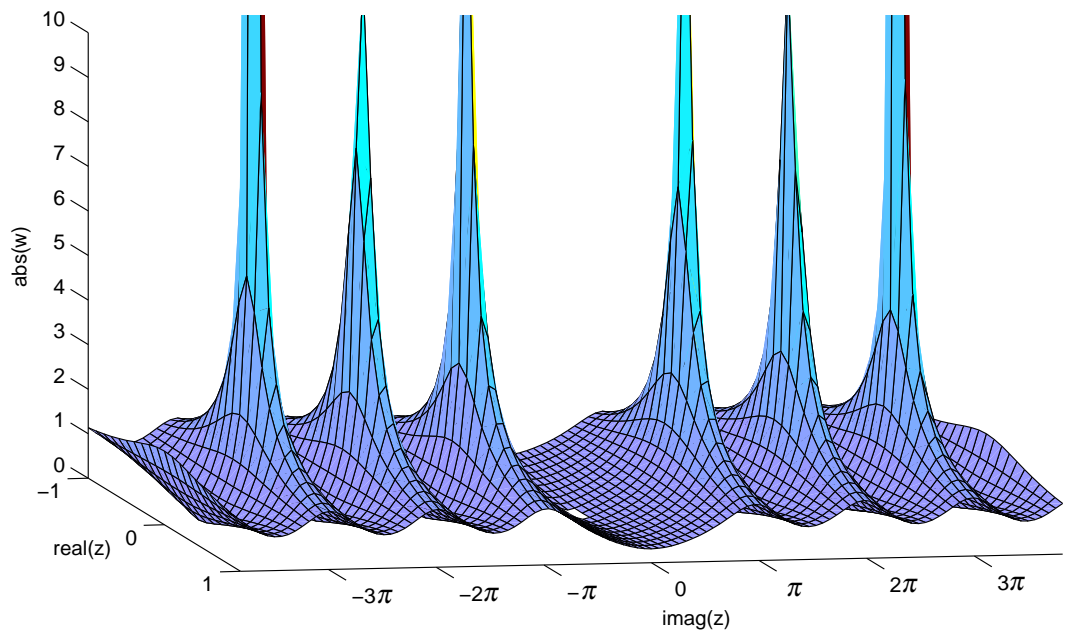
\begin{figure}[H] 
\begin{tikzpicture}
(0,0)\node   {\hspace*{-1.5cm}\includegraphics[scale=0.85]{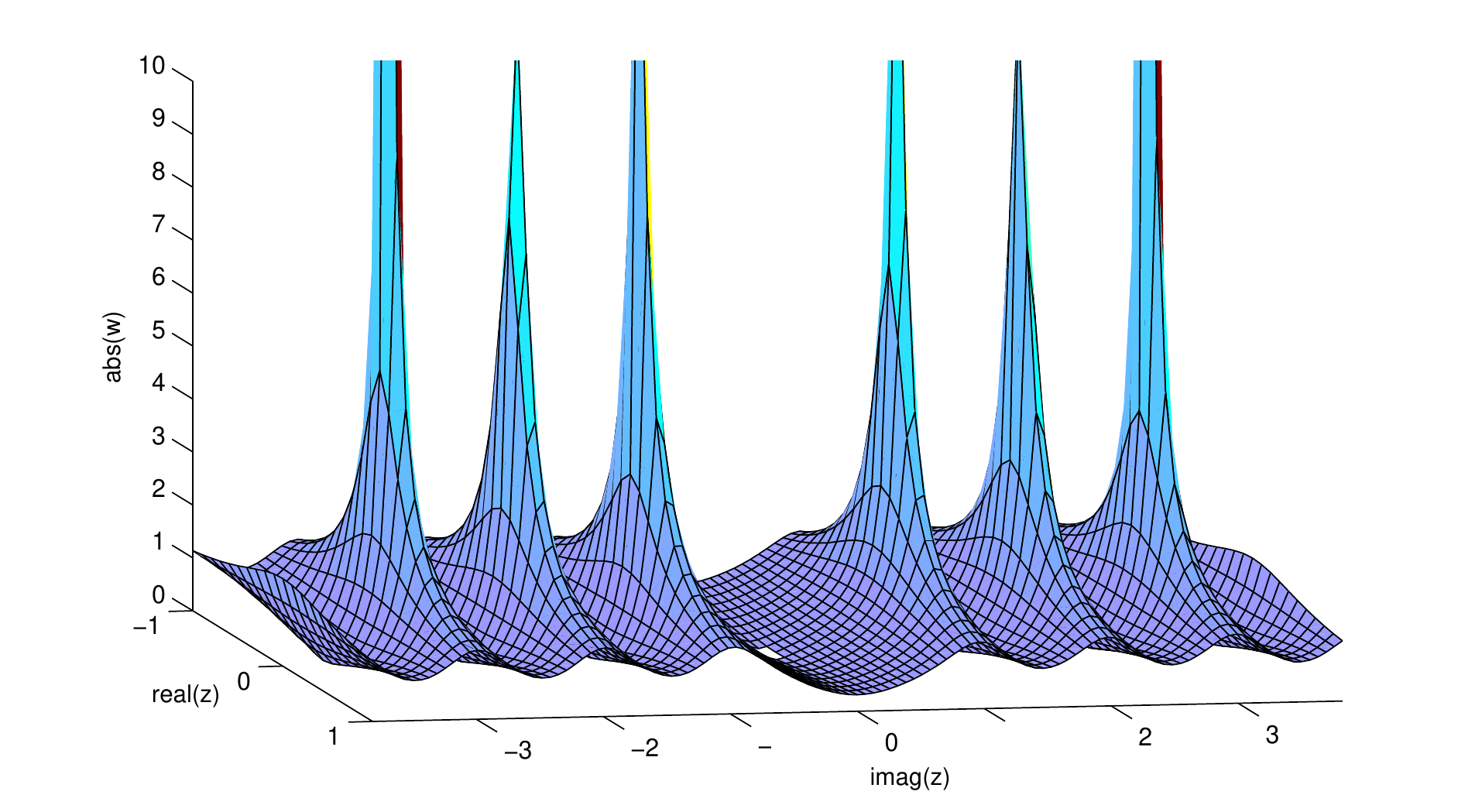}};
\draw  (-3.01,-3.86) node  {\fontsize{10}{10} $\pi$};
\draw  (-1.61,-3.83) node  {\fontsize{10}{10}  $\pi$};
\draw  (-0.32,-3.81) node  {\fontsize{10}{10}  $\pi$};
\draw  (2.38,-3.76) node {\fontsize{10}{10}  $\pi$};
\draw  (3.90,-3.71) node {\fontsize{10}{10}  $\pi$};
\draw  (5.31,-3.68) node  {\fontsize{10}{10}  $\pi$};
\end{tikzpicture}  \vspace{-5pt}
\caption{Plot of the simple poles of the Langevin function in the complex plane.  The removable singularity at the origin shows as a white square.}
\label{fig:4w}
\end{figure}

Before continuing with our discussion of the inverse Langevin function we exhibit 
Figure \ref{fig:4w} which depicts the Langevin function itself in the complex plane showing the simple poles at $z=\pm i n \pi $, $n=1,2,3, \dots$.  The removable singularity at the origin is represented by the white square.

\subsection{At the initial radius of convergence $r_1$} 
\label{sec:5.1}
\begin{figure}[H] 
\centerline{
\includegraphics[scale=0.90]{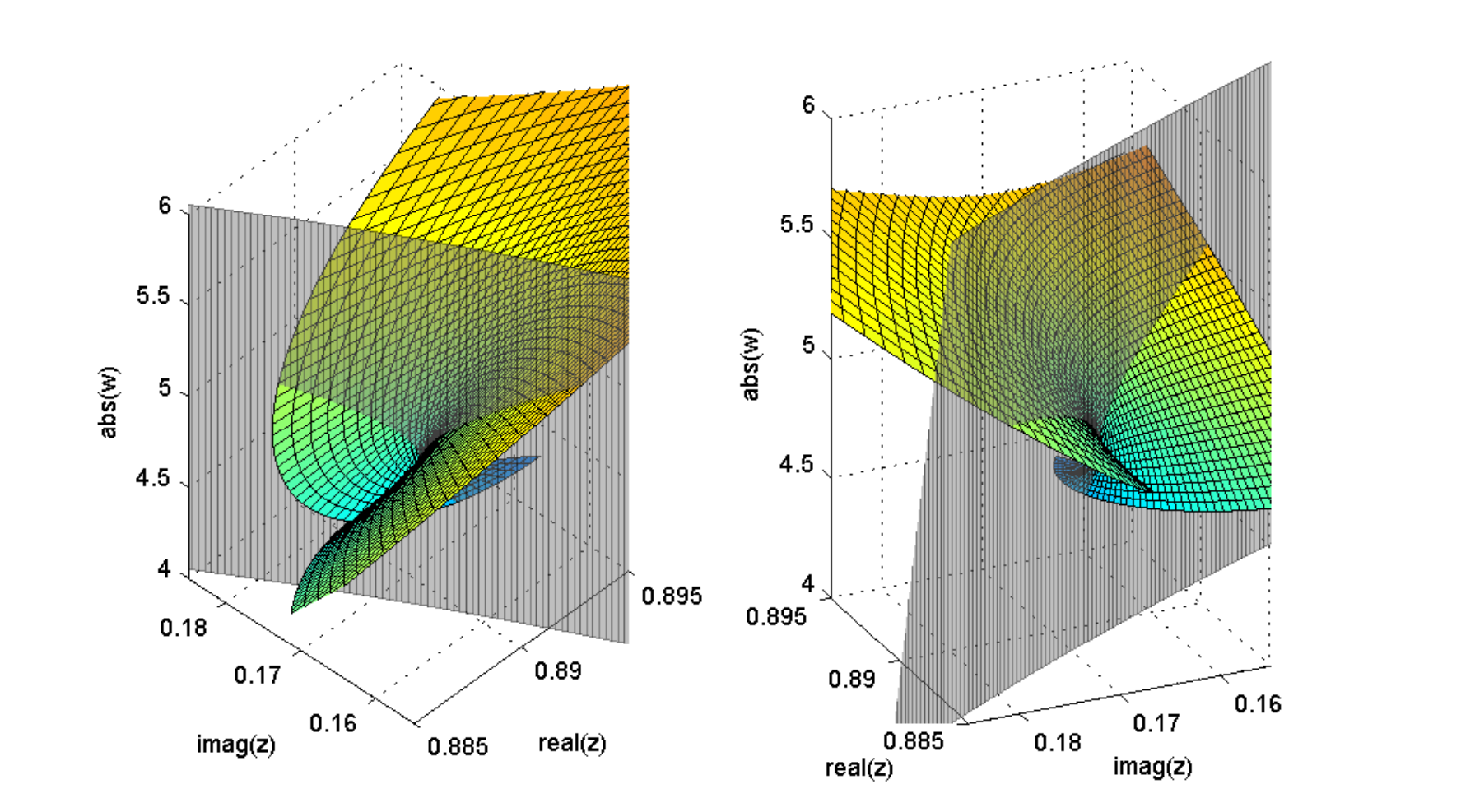}}
\caption{Plot of the inverse Langevin function $\mathscr{L}^{-1}(z)$ at the point $z=z_1=0.889+0.166i$ in the complex plane.}
\label{fig:5w}
\end{figure}

Returning to the inverse Langevin function, 
Figure \ref{fig:5w} is a three dimensional plot of this function in the complex plane specifically focusing on the singularity in the first quadrant as identified using the methods of subsection \ref{sec:4.1}. The yellow and green surface is the inverse Langevin function and the wall of grey is part of a cylinder of radius $r_1=0.905$. Figure \ref{fig:5w} clearly identifies a branch cut at $z_1 \approx 0.889+0.166i$.

Performing an extensive numerical search of the surface plotted in Figure \ref{fig:5w}  at the radius of convergence $r_1=0.9046$ we identify a branch cut singularity at
\be \label{5.1x}
z_1= 0.88924042727+0.16622770313i 
\ee
giving a more accurate estimate of the radius of convergence of $r_1=0.90464367946$.  The further branch cuts at $-z_1$ and $\pm \bar{z}_1$  were also found by this method. These are more accurate values of the positions of the singularities  given in equation (\ref{4.16x}).  These accurate values were found using MATLAB's built-in \textit{data cursor mode} which allows data points to be read directly from a plot by displaying the position of the point selected.

\begin{figure}[H] 
\centerline{
\includegraphics[scale=0.90]{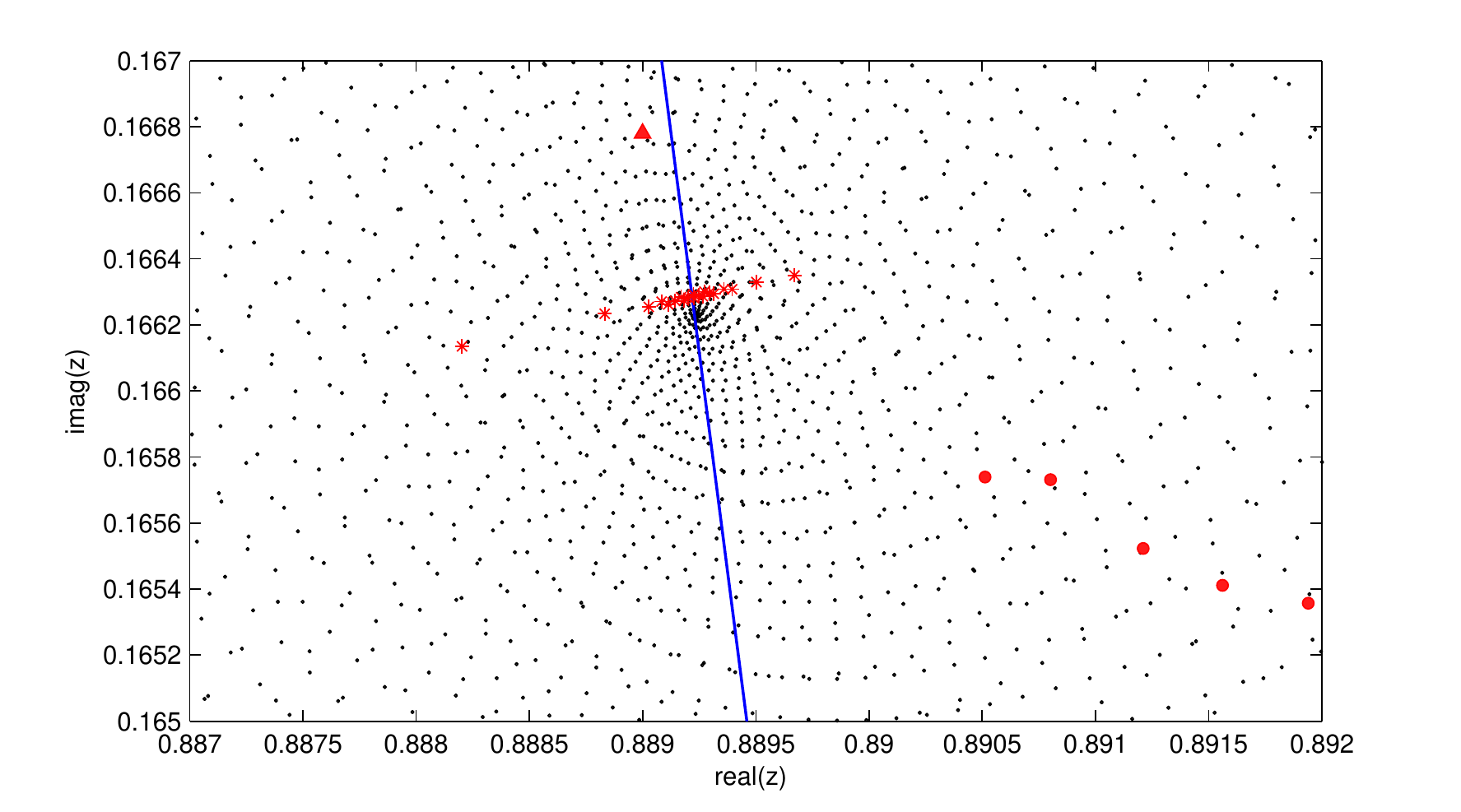}}
\caption{Singularity in the complex plane at $z=z_1$}
\label{fig:6w}
\end{figure}

Figure \ref{fig:6w} represents a two dimensional plot of the complex plane close to the point $z=z_1$ with $z_1$ given by (\ref{5.1x}).  The dense area of black dots represents the branch cut singularity. The blue curve is part of a circle at the the radius of convergence $r_1=0.9046436795$ which passes through the singularity at  $z=z_1$. The red triangle marks  the  point $z_1=0.889+0.166i$ which is the approximate value of the singularity $z_1$ found using the Domb-Sykes method.   The large red dots are the poles found using the continued fraction method and the red stars are the first 17 poles presented in Table \ref{table:1}.    Recall that the largest continued fraction we could calculate was for the Taylor series (\ref{3.3x}) expanded to 150 terms.

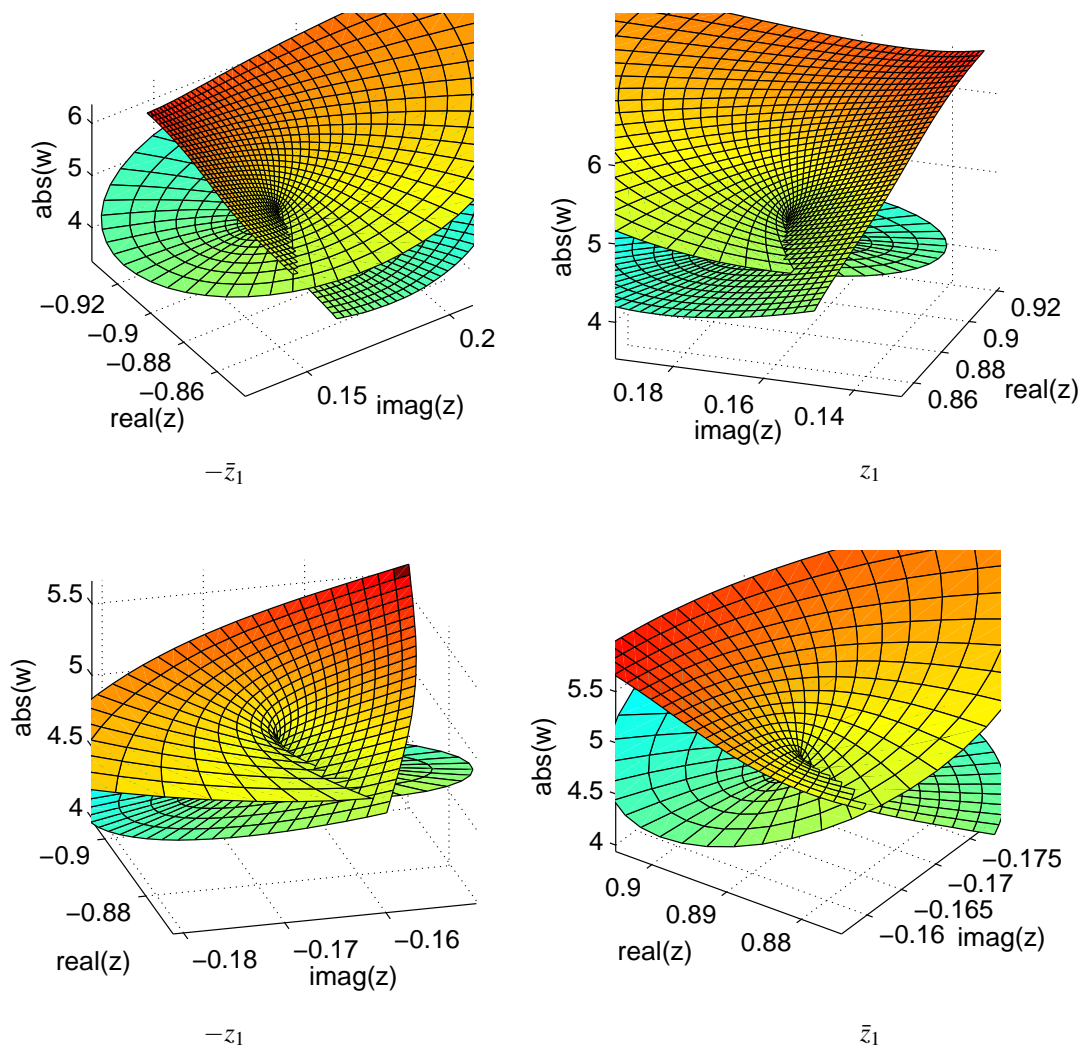
\begin{figure}[H] 
\hspace*{10mm}
\begin{tikzpicture}
(0,0)\node   {\hspace*{-1.0cm}\includegraphics[scale=0.90]{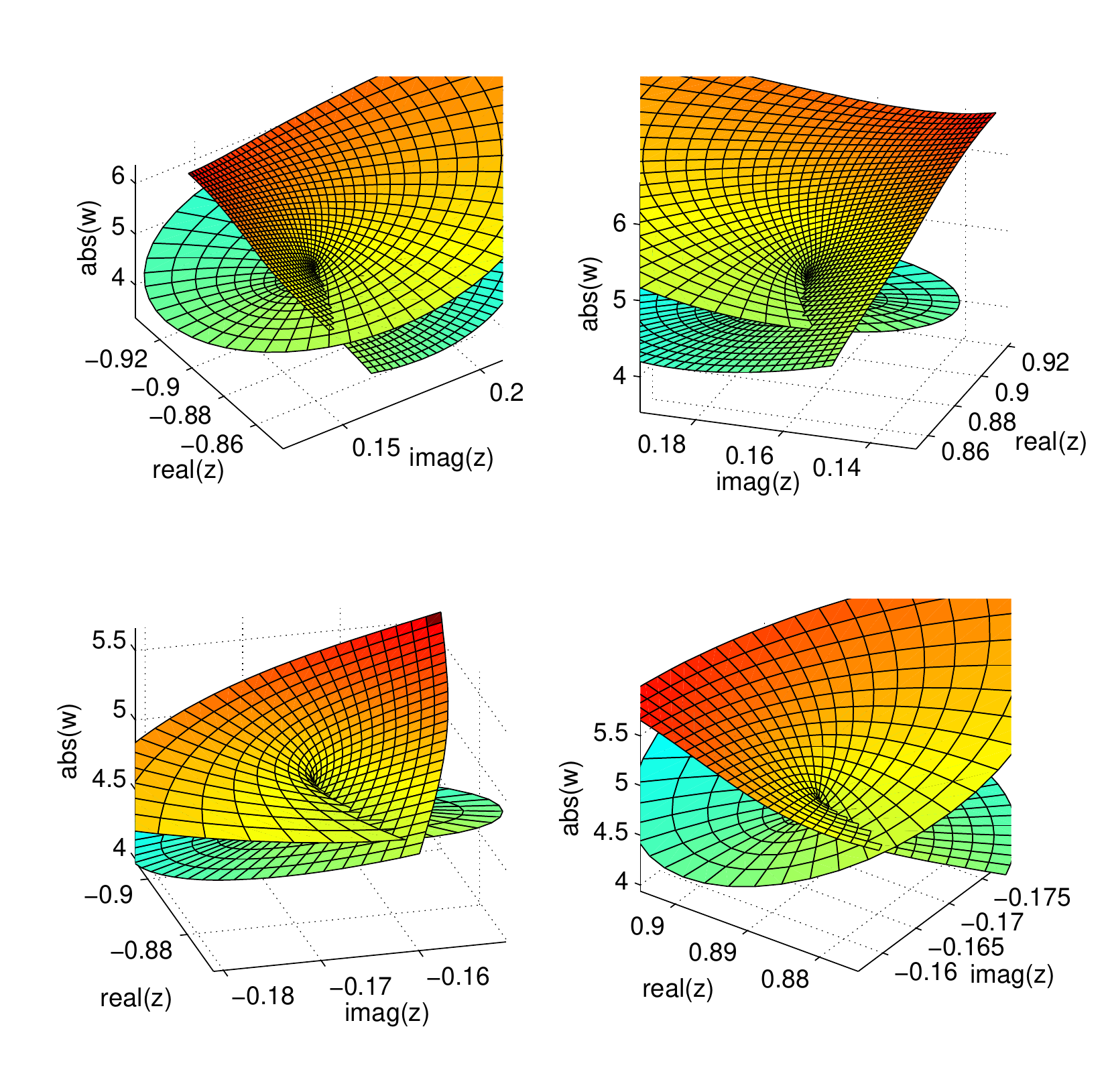}};
\draw  (-4.5,0.3) node  {\fontsize{10}{10} $-\bar{z}_1$};
\draw  (4.0,0.3) node  {\fontsize{10}{10} $z_1$};
\draw  (-4.5,-7.15) node {\fontsize{10}{10} $-z_1$};
\draw  (4.0,-7.15) node {\fontsize{10}{10} $\bar{z}_1$};
\end{tikzpicture}  \vspace{-5pt}
\caption{Plot of the four complex conjugate branch cut singularities of $\mathscr{L}^{-1}(z)$ closest to the origin} 
\label{fig:7w}
\end{figure}

Figure \ref{fig:7w} represents three dimensional plots of the four singularities nearest to the origin.  They were obtained using the same extensive numerical search methods that were used for Figure \ref{fig:5w}.  They illustrate the nature of these singularities very clearly.  They are typical square root singularities.

\subsection{At the second radius of convergence $r_2$}  
\label{sec:5.2}

\begin{figure}[H] 
\hspace*{15mm}
\begin{tikzpicture}
(0,0)\node   {\hspace*{-1.5cm}\includegraphics[scale=0.85]{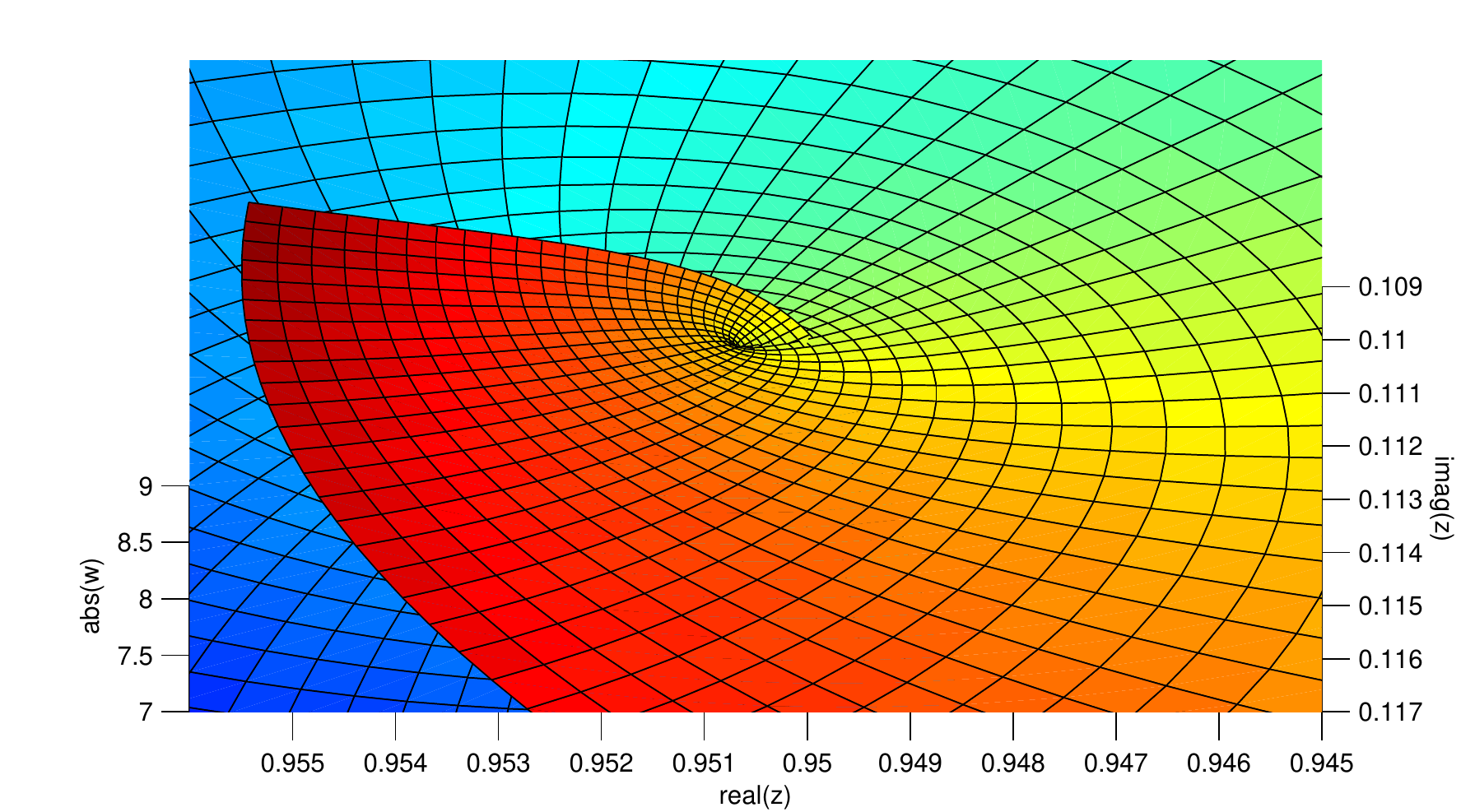}};
\end{tikzpicture}
\caption{Plot of the inverse Langevin function $\mathscr{L}^{-1}(z)$ at the point $z=z_2=0.951+0.112i$ in the complex plane.}
\label{fig:8w}
\end{figure}

Following the upper branch of the poles found using the continued fraction of the Taylor series (\ref{3.3x}) as shown in the subplot of Figure \ref{fig:3w}, we identify a new branch cut at radius of convergence $r_2=0.9573$. A graphical representation of this branch cut is shown in Figure \ref{fig:8w}. 

Using MATLAB's built-in data cursor mode as before on the surface plotted in Figure \ref{fig:8w}  at  $r_2=0.9573$ we identify a branch cut singularity at
\be \label{5.2x}
z_2=  0.95070539169+0.11225248497i, \\ 
\ee
giving a radius of convergence of $r_2=0.95730943909$ and further branch cuts at $-z_2$ and~$\pm \bar{z}_2$.

\subsection{Removal to infinity of the nearest complex  singularities by Euler's method} 
\label{sec:5.3}

Each of the four singularities $z=\pm r_1 \e^{\pm i\theta_1} $ given at (\ref{4.2x}) is a zero of the quartic  expression
\be \label{5.3x}
z^4 - 4z^2r_1^2\cos 2\theta_1+r_1^4
\ee
 and so the Euler transformation
\begin{equation}
\hat{z}=\frac{z}{\left( z^4 - 4z^2r_1^2\cos 2\theta_1+r_1^4 \right)^{1/4} }
\label{5.4x}
\end{equation}
removes each of these singularities to infinity.  This is an extension of the method of Van Dyke \cite[p. 294]{vandyke} for removing a pair of complex conjugate singularities. We do not pursue this method here.

 \section{Exact analysis of the complex singularities of the inverse Langevin function.} 
 \label{sec:6}

We extend the definition (\ref{1.2x}) of the inverse Langevin function to the complex plane by
\be \label{6.1x}
 z= \mathscr{L}(w) =  \coth w - 1/w \quad \mbox{so that} \quad w= \mathscr{L}^{-1}(z) ,
 \ee
where $w=u+iv$ and $z=x+iy$ with $u,v,x,y$ real.  

\subsection{Identifying the complex singularities of $\mathscr{L}^{-1}$} 
\label{sec:6.1}

Singularities occur when $dw/dz=0$ or $\infty$.  We find that
\be \label{6.2x}
 \frac{dz}{dw} = \frac{1}{w^2} -  \frac{1}{\sinh^2 w} 
 \ee
so that
\be \label{6.3x}
\frac{dz}{dw} =0 \implies \sinh w = \pm w. 
\ee
Since $w=0$ is a removable singularity of the righthand side of (\ref{6.2x}),  the relevant root of 
(\ref{6.3x}) closest to the origin in the first quadrant  is
\be \label{6.4x}
w_1 = u_1+ iv_1= 2.250728612 + 4.212392230i , 
\ee
which satisfies $\sinh w_1 = -w_1$ correct to 9 dp and gives rise to the branch point of $\mathscr{L}^{-1}$  at
\be \label{6.5x}
 z_1 = x_1+i y_1 = \mathscr{L}(w_1) = 0.889240427 + 0.166227703i .
 \ee
 Then the radius of convergence $r_1$ of the Taylor series  (\ref{3.2x}) of $\mathscr{L}^{-1}(z)$ is  $r_1=  |z_1| = 0.904643679 $ correct to 9 dp.  Previous estimates of $r_1$ agree quite well with this value:  $r_1\approx 0.904$ of \cite{itskov2011},   
 $r_1\approx 0.905$ predicted by 
Table \ref{table:1}, 
$r_1\approx 0.905$ from  the method of Pad\'{e} approximants,
and $r_1\approx 0.9047$ estimated at (\ref{4.15x}) by the Domb-Sykes method.

Each $w_n$ satisfies the equation
\be \label{6.6x}
 \sinh w_n -(-1)^nw_n = 0. 
 \ee
The same equation is satisfied by $-w_n$ and by $\pm \overline{w}_n$, with overbar denoting complex conjugate.  This means that every root $w_n$ of $dz/dw=0$ in the first quadrant generates another root in each of the other three quadrants.  The first 100 roots  $w_n$ of equation (\ref{6.6x}) in the first quadrant  and the corresponding branch points $z_n$ have been calculated to 15 significant figures and are exhibited in Tables \ref{table:2} and \ref{table:3}.   From these tables we can read off the values $r_1= 0.904643679457684 $ and $r_2=  0.957309439091278$ for the radius of convergence $r_1$  of the Taylor series  (\ref{3.2x}) and  the distance from the origin $r_2$ of the singularities next-nearest the origin.

From  (\ref{6.2x}) and the definition  (\ref{6.1x}) we find that
\be\label{6.7x}   \frac{dz}{dw}= 1 -\frac{2z}{w}-z^2. \ee
At the $n^{\rm th}$ branch cut we have $w=w_n$, $z=z_n$ and $ {dz}/{dw}\vert_{w_n}=0$ so that (\ref{6.7x}) reduces to
\be\label{6.8x} w_n=\frac{2z_n}{1-z_n^2}\quad\mbox{for}\quad n=1,2,\dots , \ee
which can be verified to hold to a high degree of accuracy for the $w_n$,  $z_n$ given in Tables \ref{table:2} and~\ref{table:3}. 

\begin{table} 
\centering
\setlength\tabcolsep{6pt}
\scalebox{0.78}{
\begin{tabular}{ c |c c  c  }
\hline\hline
 n  &	 $w_n = u_n+iv_n$  &  $z_n = x_n+iy_n$ &   $|z_n|$ \\\hline
 1&	2.25072861160186 $+$ 4.21239223049066$\hspace*{0.5mm}i$   & 0.889240427271280 $+$ 0.1662277031337704$\hspace*{0.5mm}i$ & 0.904643679457684 \\
3&	3.10314874582525 $+$ 10.7125373972793$\hspace*{0.5mm}i$ & 0.971651895822876 $+$ 0.0839661146033054$\hspace*{0.5mm}i$ & 0.975273148947393 \\
5&	3.55108734702208 $+$ 17.0733648531518$\hspace*{0.5mm}i$  & 0.986814304663841 $+$ 0.0554855076774800$\hspace*{0.5mm}i$ & 0.988372962727839 \\
7&	3.85880899310557 $+$ 23.3983552256513$\hspace*{0.5mm}i$ & 0.992296030564516 $+$ 0.0413207082174984$\hspace*{0.5mm}i$ & 0.993155986339352 \\
9&	4.09370492476533 $+$ 29.7081198252760$\hspace*{0.5mm}i$  & 0.994912673628922 $+$ 0.0328831365761801$\hspace*{0.5mm}i$ & 0.995455940169397 \\
11&	4.28378158777502 $+$ 36.0098660163716$\hspace*{0.5mm}i$  & 0.996372859485583 $+$ 0.0272934220634301$\hspace*{0.5mm}i$ & 0.996746610732843 \\
13&	4.44344583032427 $+$ 42.3068267176394$\hspace*{0.5mm}i$  & 0.997274223001842 $+$ 0.0233215298311909$\hspace*{0.5mm}i$ & 0.997546875899872 \\
15&	4.58110457345344 $+$ 48.6006841240946$\hspace*{0.5mm}i$  & 0.997871456587447 $+$ 0.0203554217132478$\hspace*{0.5mm}i$ & 0.998079048505216 \\
17&	4.70209646036170 $+$ 54.8924057880692$\hspace*{0.5mm}i$  & 0.998288513448522 $+$ 0.0180567357011610$\hspace*{0.5mm}i$ & 0.998451802435872 \\
19&	4.81002513746347 $+$ 61.1825901968339$\hspace*{0.5mm}i$  & 0.998591798634621 $+$ 0.0162233709022679$\hspace*{0.5mm}i$ & 0.998723574400725\\ 
21&	4.90743841652255 $+$ 67.4716286349754$\hspace*{0.5mm}i$  & 0.998819578715363 $+$ 0.0147272406524697$\hspace*{0.5mm}i$ & 0.998928146786529 \\
23&	4.99620440987113 $+$ 73.7597883468280$\hspace*{0.5mm}i$  & 0.998995207952555 $+$ 0.0134832643013575$\hspace*{0.5mm}i$ & 0.999086194443897 \\
25&	5.07773373223829 $+$ 80.0472584358892$\hspace*{0.5mm}i$  & 0.999133616637730 $+$ 0.0124327312137003$\hspace*{0.5mm}i$ & 0.999210967064024 \\
27&	5.15311770138603 $+$ 86.3341766904029$\hspace*{0.5mm}i$ & 0.999244722640744 $+$ 0.0115338272465116$\hspace*{0.5mm}i$ & 0.999311285284185 \\
29&	5.22321798924776 $+$ 92.6206460143294$\hspace*{0.5mm}i$  & 0.999335329886089 $+$ 0.0107559689029146$\hspace*{0.5mm}i$ & 0.999393212117022 \\
31&	5.28872685705572 $+$ 98.9067448937676$\hspace*{0.5mm}i$  & 0.999410236024302 $+$ 0.0100762727169390$\hspace*{0.5mm}i$ & 0.999461030326854 \\
33&	5.35020884862568 $+$ 105.192534289525$\hspace*{0.5mm}i$ & 0.999472905159809 $+$ 0.0094772770169737$\hspace*{0.5mm}i$ & 0.999517837223650 \\
35&	5.40813039638030 $+$ 111.478062307910$\hspace*{0.5mm}i$ & 0.999525890748739 $+$ 0.0089454269554641$\hspace*{0.5mm}i$ & 0.999565919257192 \\
37&	5.46288131610703 $+$ 117.763367445661$\hspace*{0.5mm}i$ & 0.999571109526232 $+$ 0.0084700403806674$\hspace*{0.5mm}i$ & 0.999606995065338 \\
39&	5.51479071941834 $+$ 124.048480894101$\hspace*{0.5mm}i$ & 0.999610023661935 $+$ 0.0080425855175257$\hspace*{0.5mm}i$ & 0.999642377346630 \\
41&	5.56413899815659 $+$ 130.333428207196$\hspace*{0.5mm}i$ & 0.999643764736604 $+$ 0.0076561660632983$\hspace*{0.5mm}i$ & 0.999673083190480 \\
43&	5.61116698984546 $+$ 136.618230530152$\hspace*{0.5mm}i$ & 0.999673219886808 $+$ 0.0073051474114691$\hspace*{0.5mm}i$ & 0.999699910842030 \\
45&	5.65608308427746 $+$ 142.902905518431$\hspace*{0.5mm}i$ & 0.999699092784112 $+$ 0.0069848808630987$\hspace*{0.5mm}i$ & 0.999723494109270 \\
47&	5.69906880245861 $+$ 149.187468034869$\hspace*{0.5mm}i$ & 0.999721947529467 $+$ 0.0066914971210470$\hspace*{0.5mm}i$ & 0.999744341572300 \\
49&	5.74028322578860 $+$ 155.471930685222$\hspace*{0.5mm}i$ & 0.999742240733079 $+$ 0.0064217495837072$\hspace*{0.5mm}i$ & 0.999762865270417 \\

51&	5.77986654860470 $+$ 161.756304234369$\hspace*{0.5mm}i$ & 0.999760345286377 $+$ 0.0061728939730420$\hspace*{0.5mm}i$ & 0.999779401981825 \\
53&	5.81794295439426 $+$ 168.040597933200$\hspace*{0.5mm}i$ & 0.999776568202020 $+$ 0.0059425948368215$\hspace*{0.5mm}i$ & 0.999794229208793 \\
55&	5.85462296453648 $+$ 174.324819777869$\hspace*{0.5mm}i$ & 0.999791164158752 $+$ 0.0057288521782264$\hspace*{0.5mm}i$ & 0.999807577325354 \\
57&	5.89000537155756 $+$ 180.608976717251$\hspace*{0.5mm}i$ & 0.999804345895960 $+$ 0.0055299433342712$\hspace*{0.5mm}i$ & 0.999819638907802 \\
59&	5.92417884209128 $+$ 186.893074820336$\hspace*{0.5mm}i$ & 0.999816292270085 $+$ 0.0053443765303652$\hspace*{0.5mm}i$ & 0.999830575972350 \\
61&	5.95722325502924 $+$ 193.177119412340$\hspace*{0.5mm}i$ & 0.999827154556475 $+$ 0.0051708534637630$\hspace*{0.5mm}i$ & 0.999840525640985 \\
63&	5.98921082568168 $+$ 199.461115186172$\hspace*{0.5mm}i$ & 0.999837061421065 $+$ 0.0050082389329232$\hspace*{0.5mm}i$ & 0.999849604614774 \\
65&	6.02020705574268 $+$ 205.745066294327$\hspace*{0.5mm}i$ & 0.999846122874077 $+$ 0.0048555360122809$\hspace*{0.5mm}i$ & 0.999857912733750 \\
67&	6.05027154047907 $+$ 212.028976425124$\hspace*{0.5mm}i$ & 0.999854433437650 $+$ 0.0047118656262774$\hspace*{0.5mm}i$ & 0.999865535830996 \\
69&	6.07945865814335 $+$ 218.312848866321$\hspace*{0.5mm}i$ & 0.999862074701456 $+$ 0.0045764496394105$\hspace*{0.5mm}i$ & 0.999872548036799 \\
71&	6.10781816164831 $+$ 224.596686558490$\hspace*{0.5mm}i$ & 0.999869117398106 $+$ 0.0044485967760410$\hspace*{0.5mm}i$ & 0.999879013651023 \\
73&	6.13539568867313 $+$ 230.880492140035$\hspace*{0.5mm}i$ & 0.999875623098991 $+$ 0.0043276908325902$\hspace*{0.5mm}i$ & 0.999884988673966 \\
75&	6.16223320333355 $+$ 237.164267985343$\hspace*{0.5mm}i$ & 0.999881645608026 $+$ 0.0042131807582873$\hspace*{0.5mm}i$ & 0.999890522065250 \\
77&	6.18836938014610 $+$ 243.448016237264$\hspace*{0.5mm}i$ & 0.999887232113431 $+$ 0.0041045722678650$\hspace*{0.5mm}i$ & 0.999895656784726 \\
79&	6.21383993910375 $+$ 249.731738834892$\hspace*{0.5mm}i$ & 0.999892424144466 $+$ 0.0040014207171476$\hspace*{0.5mm}i$ & 0.999900430657599 \\
81&	6.23867793914730 $+$ 256.015437537410$\hspace*{0.5mm}i$ & 0.999897258370095 $+$ 0.0039033250251526$\hspace*{0.5mm}i$ & 0.999904877096959 \\
83&	6.26291403608140 $+$ 262.299113944652$\hspace*{0.5mm}i$ & 0.999901767268770 $+$ 0.0038099224676771$\hspace*{0.5mm}i$ & 0.999909025710048 \\
85&	6.28657670998225 $+$ 268.582769514888$\hspace*{0.5mm}i$ & 0.999905979692632 $+$ 0.0037208842000094$\hspace*{0.5mm}i$ & 0.999912902809196 \\
87&	6.30969246632743 $+$ 274.866405580268$\hspace*{0.5mm}i$ & 0.999909921344788 $+$ 0.0036359113923740$\hspace*{0.5mm}i$ & 0.999916531844230 \\
89&	6.33228601440944 $+$ 281.150023360279$\hspace*{0.5mm}i$ & 0.999913615184637 $+$ 0.0035547318824756$\hspace*{0.5mm}i$ & 0.999919933769883 \\
91&	6.35438042604379 $+$ 287.433623973499$\hspace*{0.5mm}i$ & 0.999917081773447 $+$ 0.0034770972661930$\hspace*{0.5mm}i$ & 0.999923127359161 \\
93&	6.37599727712690 $+$ 293.717208447911$\hspace*{0.5mm}i$ & 0.999920339570038 $+$ 0.0034027803609515$\hspace*{0.5mm}i$ & 0.999926129471595 \\
95&	6.39715677422076 $+$ 300.000777729961$\hspace*{0.5mm}i$ & 0.999923405184657 $+$ 0.0033315729872513$\hspace*{0.5mm}i$ & 0.999928955283640 \\
97&	6.41787786802549 $+$ 306.284332692552$\hspace*{0.5mm}i$ & 0.999926293597703 $+$ 0.0032632840227542$\hspace*{0.5mm}i$ & 0.999931618487311 \\
99&	6.43817835533659 $+$ 312.567874142105$\hspace*{0.5mm}i$ & 0.999929018348732 $+$ 0.0031977376906519$\hspace*{0.5mm}i$ & 0.999934131461767 \\
	[1ex] 
\hline 
\end{tabular}}
\caption{Solutions $w_n$ of $\sinh w=-w$ in the first quadrant and the corresponding branch points $z_n= \mathscr{L}(w_n)$.}
\label{table:2}
\end{table}

\begin{table} 
\centering
\setlength\tabcolsep{6pt}
\scalebox{0.78}{
\begin{tabular}{ c |c  c c  }
\hline\hline
 n &	 $w_n = u_n+iv_n$  &  $z_n= x_n+iy_n$ &    $|z_n|$ \\	 \hline
2&	2.76867828298732 $+$ 7.49767627777639$\hspace*{0.5mm}i$ & 0.9507053916921612 $+$ 0.1122524849645265$\hspace*{0.5mm}i$ & 0.957309439091278 \\
4&	3.35220988485350 $+$ 13.8999597139765$\hspace*{0.5mm}i$ & 0.9814249275768375 $+$ 0.0668711449824057$\hspace*{0.5mm}i$ & 0.983700482108482 \\
6&	3.71676767975250 $+$ 20.2385177078300$\hspace*{0.5mm}i$ & 0.9901175152938447 $+$ 0.0473785974133679$\hspace*{0.5mm}i$ & 0.991250435351488 \\
8&	3.98314164033996 $+$ 26.5545472654916$\hspace*{0.5mm}i$ & 0.9938124512702532 $+$ 0.0366260858125753$\hspace*{0.5mm}i$ & 0.994487133381694 \\
10&	4.19325147043121 $+$ 32.8597410050699$\hspace*{0.5mm}i$ & 0.9957375998196626 $+$ 0.0298302646937089$\hspace*{0.5mm}i$ & 0.996184326511073 \\
12&	4.36679511767062 $+$ 39.1588165200650$\hspace*{0.5mm}i$ & 0.9968730165101870 $+$ 0.0251523955760530$\hspace*{0.5mm}i$ & 0.997190279760755 \\
14&	4.51464044948130 $+$ 45.4540714643551$\hspace*{0.5mm}i$ & 0.9976012288775898 $+$ 0.0217381656303724$\hspace*{0.5mm}i$ & 0.997838042822106 \\
16&	4.64342795705190 $+$ 51.7467683028218$\hspace*{0.5mm}i$ & 0.9980974683847858 $+$ 0.0191375308460208$\hspace*{0.5mm}i$ & 0.998280923128856 \\
18&	4.75751511808162 $+$ 58.0376620590943$\hspace*{0.5mm}i$ & 0.9984515279054803 $+$ 0.0170911687514174$\hspace*{0.5mm}i$ & 0.998597797727432 \\
20&	4.85991664789710 $+$ 64.3272337132856$\hspace*{0.5mm}i$ & 0.9987134143666453 $+$ 0.0154392353115651$\hspace*{0.5mm}i$ & 0.998832745770225 \\
22&	4.95280535741894 $+$ 70.6158050613296$\hspace*{0.5mm}i$ & 0.9989128315092538 $+$ 0.0140778854034248$\hspace*{0.5mm}i$ & 0.999012027861160 \\
24&	5.03779919329181 $+$ 76.9036000092884$\hspace*{0.5mm}i$ & 0.9990683548450882 $+$ 0.0129367469874774$\hspace*{0.5mm}i$ & 0.999152109078237 \\
26&	5.11613546596693 $+$ 83.1907794378375$\hspace*{0.5mm}i$ & 0.9991920999701078 $+$ 0.0119664511932357$\hspace*{0.5mm}i$ & 0.999263753268793 \\
28&	5.18878162856979 $+$ 89.4774620851630$\hspace*{0.5mm}i$ & 0.9992922511676224 $+$ 0.0111313467283301$\hspace*{0.5mm}i$ & 0.999354246563069 \\
30&	5.25650846760145 $+$ 95.7637376020254$\hspace*{0.5mm}i$ & 0.9993745037000536 $+$ 0.0104050483334986$\hspace*{0.5mm}i$ & 0.999428668628508 \\
32&	5.31994004501781 $+$ 102.049675012746$\hspace*{0.5mm}i$ & 0.9994429230481451 $+$ 0.0097676120844439$\hspace*{0.5mm}i$ & 0.999490651620540 \\
34&	5.37958872776665 $+$ 108.335328370326$\hspace*{0.5mm}i$ & 0.9995004761610887 $+$ 0.0092036840636058$\hspace*{0.5mm}i$ & 0.999542850330383 \\
36&	5.43588034897744 $+$ 114.620740638271$\hspace*{0.5mm}i$ & 0.9995493706262943 $+$ 0.0087012525134125$\hspace*{0.5mm}i$ & 0.999587242873138 \\
38&	5.48917265997610 $+$ 120.905946417967$\hspace*{0.5mm}i$ & 0.9995912773595068 $+$ 0.0082507858904696$\hspace*{0.5mm}i$ & 0.999625328431118 \\
40&	5.53976911179991 $+$ 127.190973904692$\hspace*{0.5mm}i$ & 0.9996274804707284 $+$ 0.0078446244699579$\hspace*{0.5mm}i$ & 0.999658260529732 \\
42&	5.58792931680847 $+$ 133.475846316267$\hspace*{0.5mm}i$ & 0.9996589803554569 $+$ 0.0074765425850377$\hspace*{0.5mm}i$ & 0.999686938843525 \\
44&	5.63387710617850 $+$ 139.760582953693$\hspace*{0.5mm}i$ & 0.9996865660131676 $+$ 0.0071414281811005$\hspace*{0.5mm}i$ & 0.999712073681050 \\
46&	5.67780681724206 $+$ 146.045200000247$\hspace*{0.5mm}i$ & 0.9997108666807187 $+$ 0.0068350445868388$\hspace*{0.5mm}i$ & 0.999734232080716 \\
48&	5.71988825776711 $+$ 152.329711131562$\hspace*{0.5mm}i$ & 0.9997323892895176 $+$ 0.0065538509093109$\hspace*{0.5mm}i$ & 0.999753871288464 \\
50&	5.76027066783195 $+$ 158.614127987106$\hspace*{0.5mm}i$ & 0.9997515460351737 $+$ 0.0062948648906245$\hspace*{0.5mm}i$ & 0.999771363424514 \\
52&	5.79908591278778 $+$ 164.898460538558$\hspace*{0.5mm}i$ & 0.9997686749398286 $+$ 0.0060555569619736$\hspace*{0.5mm}i$ & 0.999787013898991 \\
54&	5.83645107971304 $+$ 171.182717380588$\hspace*{0.5mm}i$ & 0.9997840553752877 $+$ 0.0058337675202634$\hspace*{0.5mm}i$ & 0.999801075327556 \\
56&	5.87247060628390 $+$ 177.466905962516$\hspace*{0.5mm}i$ & 0.9997979199133771 $+$ 0.0056276416996543$\hspace*{0.5mm}i$ & 0.999813758164097 \\
58&	5.90723803960322 $+$ 183.751032774485$\hspace*{0.5mm}i$ & 0.9998104634661614 $+$ 0.0054355774695200$\hspace*{0.5mm}i$ & 0.999825238908704 \\
60&	5.94083749958616 $+$ 190.035103498259$\hspace*{0.5mm}i$ & 0.9998218504033833 $+$ 0.0052561839878047$\hspace*{0.5mm}i$ & 0.999835666504330 \\
62&	5.97334490452427 $+$ 196.319123130293$\hspace*{0.5mm}i$ & 0.9998322201440941 $+$ 0.0050882479216131$\hspace*{0.5mm}i$ & 0.999845167366018 \\
64&	6.00482900375061 $+$ 202.603096082861$\hspace*{0.5mm}i$ & 0.9998416915859598 $+$ 0.0049307060121844$\hspace*{0.5mm}i$ & 0.999853849367623 \\
66&	6.03535225272971 $+$ 208.887026267693$\hspace*{0.5mm}i$ & 0.9998503666409769 $+$ 0.0047826225743309$\hspace*{0.5mm}i$ & 0.999861805026467 \\
68&	6.06497155857294 $+$ 215.170917165564$\hspace*{0.5mm}i$ & 0.9998583330782893 $+$ 0.0046431709252659$\hspace*{0.5mm}i$ & 0.999869114065604 \\
70&	6.09373891834152 $+$ 221.454771884529$\hspace*{0.5mm}i$ & 0.9998656668253923 $+$ 0.0045116179650518$\hspace*{0.5mm}i$ & 0.999875845489253 \\
72&	6.12170196812237 $+$ 227.738593208909$\hspace*{0.5mm}i$ & 0.9998724338427762 $+$ 0.0043873113019690$\hspace*{0.5mm}i$ & 0.999882059274561 \\
74&	6.14890445743776 $+$ 234.022383640710$\hspace*{0.5mm}i$ & 0.9998786916602354 $+$ 0.0042696684459880$\hspace*{0.5mm}i$ & 0.999887807758862 \\
76&	6.17538666085092 $+$ 240.306145434805$\hspace*{0.5mm}i$ & 0.9998844906430244 $+$ 0.0041581676929462$\hspace*{0.5mm}i$ & 0.999893136783638 \\
78&	6.20118573648736 $+$ 246.589880628955$\hspace*{0.5mm}i$ & 0.9998898750409366 $+$ 0.0040523403987275$\hspace*{0.5mm}i$ & 0.999898086642878 \\
80&	6.22633603948128 $+$ 252.873591069537$\hspace*{0.5mm}i$ & 0.9998948838619153 $+$ 0.0039517644023399$\hspace*{0.5mm}i$ & 0.999902692873224 \\
82&	6.25086939698059 $+$ 259.157278433673$\hspace*{0.5mm}i$ & 0.9998995516030284 $+$ 0.0038560584034185$\hspace*{0.5mm}i$ & 0.999906986915457 \\
84&	6.27481535023279 $+$ 265.440944248347$\hspace*{0.5mm}i$ & 0.9999039088648717 $+$ 0.0037648771364091$\hspace*{0.5mm}i$ & 0.999910996670755 \\
86&	6.29820136836983 $+$ 271.724589906972$\hspace*{0.5mm}i$ & 0.9999079828702210 $+$ 0.0036779072127901$\hspace*{0.5mm}i$ & 0.999914746970491 \\
88&	6.32105303777144 $+$ 278.008216683796$\hspace*{0.5mm}i$ & 0.9999117979036525 $+$ 0.0035948635258953$\hspace*{0.5mm}i$ & 0.999918259974626 \\
90&	6.34339423028012 $+$ 284.291825746477$\hspace*{0.5mm}i$ & 0.9999153756856314 $+$ 0.0035154861314920$\hspace*{0.5mm}i$ & 0.999921555510870 \\
92&	6.36524725303997 $+$ 290.575418167091$\hspace*{0.5mm}i$ & 0.9999187356920239 $+$ 0.0034395375322614$\hspace*{0.5mm}i$ & 0.999924651364477 \\
94&	6.38663298231693 $+$ 296.858994931788$\hspace*{0.5mm}i$ & 0.9999218954279676 $+$ 0.0033668003064646$\hspace*{0.5mm}i$ & 0.999927563526761 \\
96&	6.40757098331232 $+$ 303.142556949306$\hspace*{0.5mm}i$ & 0.9999248706634182 $+$ 0.0032970750309580$\hspace*{0.5mm}i$ & 0.999930306408908 \\
98&	6.42807961769295 $+$ 309.426105058477$\hspace*{0.5mm}i$ & 0.9999276756363949 $+$ 0.0032301784568029$\hspace*{0.5mm}i$ & 0.999932893026560 \\
100& 6.44817614031860 $+$ 315.709640034877$\hspace*{0.5mm}i$ & 0.9999303232289011 $+$ 0.0031659419023442$\hspace*{0.5mm}i$ & 0.999935335159621 \\
	[1ex] 
\hline 
\end{tabular}}
\caption{Solutions $w_n$ of $\sinh w=w$ in the first quadrant and the corresponding branch points $z_n= \mathscr{L}(w_n)$.
}
\label{table:3}
\end{table}

\subsection{Power series for $w$ close to the first singularity $z=z_1$} 
\label{sec:6.2}
 
 We expand $z=\sL(w)$ as a power series about the point $w_1$ which is possible since $\sL$ is analytic everywhere in the complex plane except at its poles $w= \pm n\pi i,\; n=1,2,\ldots$. 
 \be \label{6.9x}
 z=\sL(w_1+(w-w_1))=\sL(w_1)+(w-w_1)\frac{d\sL}{d w}\Big|_{w=w_1}+
 \frac12(w-w_1)^2\frac{d^2\sL}{d w^2}\Big|_{w=w_1}+\cdots
 \ee
 At $w=w_1$,  we have $z=z_1$,  ${d\sL}/{d w}=0$ and, by differentiating (\ref{6.7x}),  
 ${d^2\sL}/{d w^2} =  2{z_1}/{w_1^2}$,
 so the series (\ref{6.9x}) becomes
 \be\label{6.10x} z-z_1=   \frac{z_1}{w_1^2}(w-w_1)^2+ O\left((w-w_1)^3 \right).
 \ee
Taking $w-w_1$ small in (\ref{6.10x}) we see that $z-z_1$ and $(w-w_1)^2$ must balance.  Therefore the series for $w$ must take the form
\be \label{6.11x}
 w-w_1=\frac{w_1}{\sqrt z_1}(z-z_1)^{1/2} + \cdots
 \ee
 and continues as a power series in $(z-z_1)^{1/2}$.
The appearance of the exponent $1/2$ in (\ref{6.11x}) perhaps explains the exponent $\al\approx 1/2$ in the Domb-Sykes plots, see (\ref{4.15x}), and the typical square root nature of the plots in Figure \ref{fig:7w}.

\begin{figure}[H] 
\centerline{
\includegraphics[scale=0.90]{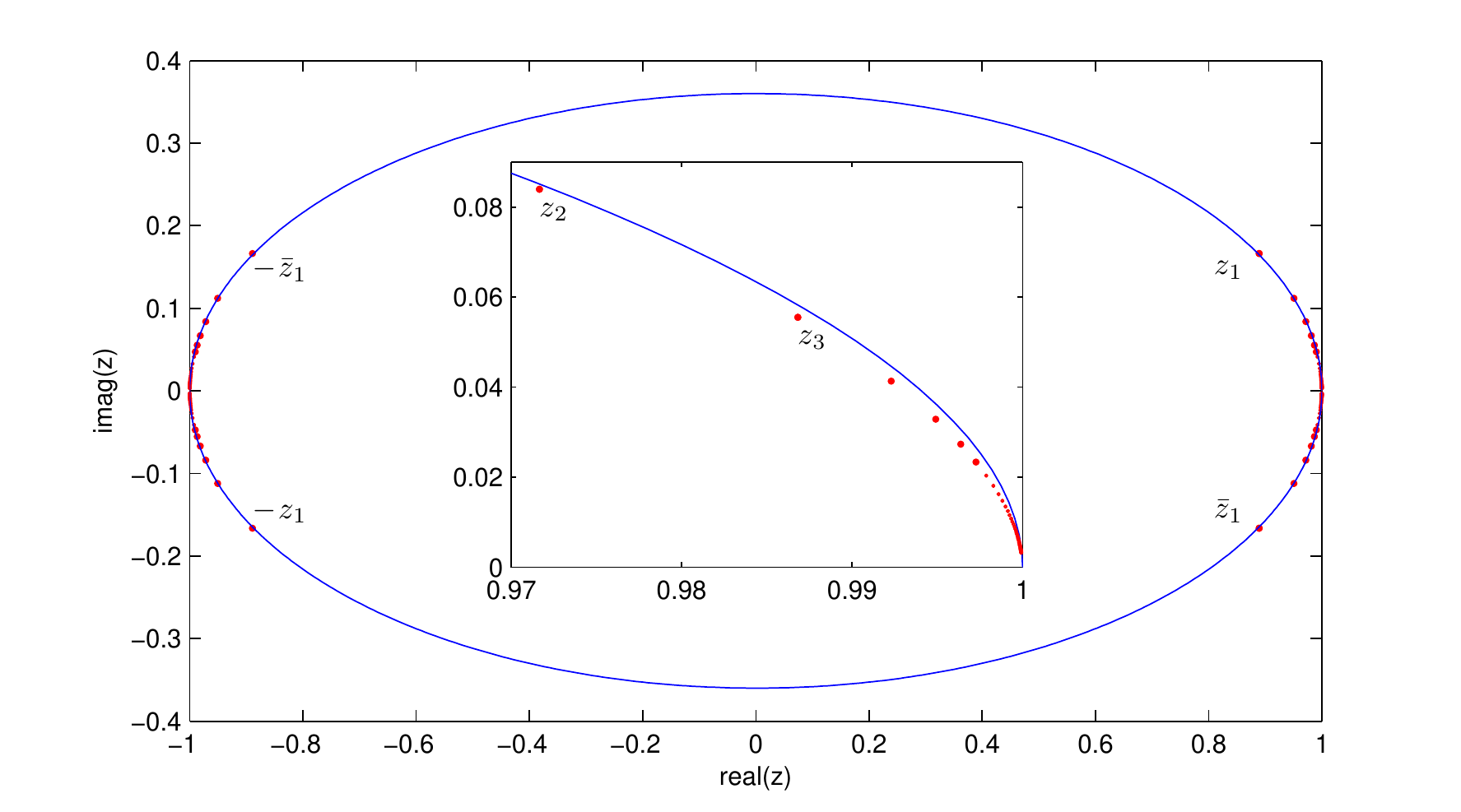}}
\caption{The branch cut singularities lie approximately on the ellipse $x^2+y^2/(0.36)^2=1$.  The inset shows the accumulation point of singularities at $z=1$.}
\label{fig:9w}
\end{figure}

 On writing $w_n=u_n+i v_n$ we observe the following approximations from Tables  \ref{table:2} and~\ref{table:3}
\be \label{6.12x} v_n\approx (n+\tfrac12)\pi\quad\mbox{and}\quad u_n\approx \sinh^{-1}v_n. \ee    
We further observe empirically that the branch cuts at $z_n=x_n+ i y_n$ given in Tables \ref{table:2} and~\ref{table:3} lie approximately on the ellipse
\be \label{6.13x}
x^2+y^2/(0.36)^2 = 1.
\ee
This is illustrated in Figure \ref{fig:9w}.

\section{Conclusions} 
\label{sec:7}

The inverse Langevin function has been used extensively in the modern literature, spanning nearly 80 years, to model the limited-stretch elasticity of rubber and rubber-like materials beginning with the original single chain model of Kuhn \& Gr\"un \cite{kuhn}. 
We give a Taylor series for the inverse Langevin function and note that its only real singularities  are two simple poles at $x=\pm1$, each with residue $-1$.   We have seen that the effects of these poles may be removed either  multiplicatively or additively but it is evident there remain complex singularities.     

In Section \ref{sec:4}  we extended  the methods of Hunter \& Guerrieri \cite{hunter} and Mercer \& Roberts \cite{mercer} in the two-singularity case to the present situation of four complex singularities equidistant from the origin and of equal strength and then developed an algorithm for estimating the four complex  singularities with the smallest radius of convergence. 
Using these algorithms and excluding the two outliers from the sequence $2m=264..280$ presented in Table \ref{table:1}, we find that the average values of $r= 0.9047$ and $\cos2\theta=0.9324$ obtained from eq.(\ref{4.7x}) are  correct to 0.01\%, demonstrating excellent agreement between the new algorithm and the exact values found in Section \ref{sec:6}. 
From the Domb-Sykes \cite{domb} plot the radius of convergence is estimated to be $r= 0.9047$  and  the order of the singularity is estimated to be  $\alpha=0.50$, correct to 2 dp.   Also in Section \ref{sec:4}  we used the method of Pad\'{e} approximants to show that the positions of the poles tend to imply a radius of convergence of $r_1\approx 0.905$, see Figure \ref{fig:3w}.
Itskov et al. \cite{itskov2011} had earlier used the Taylor series (\ref{3.2x}) to estimate its radius of convergence to be $r_1 \approx0.904$.   These estimates of the radius of convergence  compare well with the exact value $r_1 \approx 0.9046$, correct to 4dp, found in  Section \ref{sec:6}.

 As an illustrative example of the complex singularities, in Section \ref{sec:5} we presented a graphical representation of the four singularities nearest the origin which points to these complex singularities being of a square root nature, see Figure \ref{fig:7w}.
These methods were then used to  discuss and illustrate the branch cut singularities which are the next-nearest to the origin. 

An exact analysis of the complex singularities of the inverse Langevin function was given in Section \ref{sec:6}.    We found that $\mathscr{L}^{-1}(x)$ has an infinity of complex singularities.   The complex singularities have been identified as square root branch points and the first 100 are given in Tables \ref{table:2} and \ref{table:3} correct to 15 significant figures.
From these tables we can read off the values $r_1= 0.904643679457684 $ and $r_2=  0.957309439091278$ for the radius of convergence $r_1$  of the Taylor series  (\ref{3.2x}) and  the distance from the origin $r_2$ of the singularities next-nearest the origin.

\clearpage
\bibliographystyle{plain}
\bibliography{BIBLIOGRAPHYa}

\appendix
\renewcommand{\theequation}{\Alph{section}.\arabic{equation}}

\section{Power series expansion for h(x) defined by (\ref{3.7x})}

\allowdisplaybreaks
\begin{align}  \label{A1}
\hspace*{-8mm}h(x)=&\,
1.000000000-0.200000000x^{2}-0.302857143x^{4}-0.241142857x^{6} \\  
&-0.128133581x^{8}-0.002754320x^{10}+0.112823652x^{12}+0.202302872x^{14} \nonumber \\
&+0.252957871x^{16}+0.255767493x^{18}+0.206400668x^{20}+0.106427839x^{22} \nonumber \\
&-0.035638662x^{24}-0.203814671x^{26}-0.374937806x^{28}-0.520051694x^{30} \nonumber \\  
&-0.607058580x^{32}-0.604674268x^{34}-0.487519715x^{36}-0.241938122x^{38} \nonumber \\  
&+0.128142774x^{40}+0.596240341x^{42}+1.11037244x^{44}+1.59335189x^{46} \nonumber \\  
&+1.94743654x^{48}+2.06412582x^{50}+1.83930767x^{52}+1.19311006x^{54} \nonumber \\  
&+0.092746229x^{56}-1.42453659x^{58}-3.23242249x^{60}-5.10527925x^{62} \nonumber \\  
&-6.72188993x^{64}-7.68786819x^{66}-7.57992954x^{68}-6.01247167x^{70} \nonumber \\  
&-2.72302819x^{72}+2.33166258x^{74}+8.88247800x^{76}+16.2760008x^{78} \nonumber \\  
&+23.4457067x^{80}+28.9488707x^{82}+31.0887121x^{84}+28.1332646x^{86} \nonumber \\  
&+18.6289734x^{88}+1.78834329x^{90}-22.0915299x^{92}-51.2463614x^{94} \nonumber \\  
&-82.1856215x^{96}-109.657418x^{98}-126.936675x^{100}-126.515484x^{102} \nonumber \\  
&-101.230289x^{104}-45.7910264x^{106}+41.4148208x^{108}+156.476988x^{110} \nonumber \\  
&+288.416518x^{112}+418.365849x^{114}+519.941655x^{116}+561.243625x^{118} \nonumber \\  
&+508.779920x^{120}+333.372235x^{122}+17.7285206x^{124}-435.089332x^{126} \nonumber \\  
&-993.583683x^{128}-1591.88222x^{130}-2128.03573x^{132}-2468.71682x^{134} \nonumber \\  
&-2462.15815x^{136}-1960.33797x^{138}-850.028856x^{140}+909.598564x^{142} \nonumber \\  
&+3247.43267x^{144}+5945.95697x^{146}+8620.94561x^{148}+10725.5600x^{150} \nonumber \\  
&+11588.7216x^{152}+10495.1411x^{154}+6809.31069x^{156}+138.002694x^{158} \nonumber \\  
&-9484.30277x^{160}-21414.1097x^{162}-34259.9150x^{164}-45832.1674x^{166} \nonumber \\  
&-53231.6417x^{168}-53119.8507x^{170}-42197.9104x^{172}-17890.0779x^{174} \nonumber \\  
&+20814.7257x^{176}+72462.7741x^{178}+132332.352x^{180}+191933.101x^{182} \nonumber \\  
&+239044.368x^{184}+258526.027x^{186}+234083.723x^{188}+151058.955x^{190} \nonumber \\  
&+138.520758x^{192}-218355.931x^{194}-490207.857x^{196}-783953.594x^{198} \nonumber \\  
&-1049528.05x^{200}-1220084.59x^{202}-1218051.56x^{204}-966106.693x^{206} \nonumber \\ 
&-403058.159x^{208}+496382.913x^{210}+1700206.49x^{212}+3099711.64x^{214} \nonumber \\  
&+4496983.88x^{216}+5604915.71x^{218}+6065463.63x^{220}+5490700.59x^{222} \nonumber \\  
&+3528641.83x^{224}-48340.4005x^{226}-5240311.82x^{228}-11716162.1x^{230} \nonumber \\  
&-18730611.2x^{232}-25088247.2x^{234}-29183447.1x^{236}-29142674.8x^{238} \nonumber \\  
&-23086790.5x^{240}-9514471.49x^{242}+12216497.8x^{244}+41364045.0x^{246} \nonumber \\  
&+75320486.9x^{248}+109294073.x^{250}+136293596.x^{252}+147558398.x^{254} \nonumber \\  
&+133549963.x^{256}+85559226.7x^{258}-2117991.14x^{260}-129623899.x^{262} \nonumber \\  
&-288950873.x^{264}-461841775.x^{266}-618836155.x^{268}-720187300.x^{270} \nonumber \\  
&-719324164.x^{272}-569318735.x^{274}-232412338.x^{276}+307951890.x^{278} \nonumber \\  
&+0.103391594e10x^{280}+0.188099576e10x^{282}+0.272985768e10x^{284}+0.340562715e10x^{286} \nonumber \\  
&+0.368836021e10x^{288}+0.333768440e10x^{290}+0.213308037e10x^{292}-71274982.4x^{294} \nonumber \\  
&-0.328169584e10x^{296}-0.729899021e10x^{298}-0.116643883e11x^{300}-0.156341281e11x^{302} \nonumber \\  
&-0.182012750e11x^{304}-0.181822064e11x^{306}-0.143799103e11x^{308}-0.582598386e10x^{310} \nonumber \\  
&+0.791220705e10x^{312}+0.263925558e11x^{314}+0.479829063e11x^{316}+0.696458232e11x^{318} \nonumber \\  
&+0.869147211e11x^{320}+0.941554975e11x^{322}+0.851924977e11x^{324}+0.543388085e11x^{326} \nonumber \\  
&-0.219436431e10x^{328}-0.846249706e11x^{330}-0.187888503e12x^{332}-0.300224905e12x^{334} \nonumber \\  
&-0.402497064e12x^{336}-0.468724442e12x^{338}-0.468288971e12x^{340}-0.370136642e12x^{342} \nonumber \\  
&-0.149034695e12x^{344}+0.206455043e12x^{346}+0.685141918e12x^{348}+0.124494613e13x^{350} \nonumber \\  
&+0.180719994e13x^{352}+0.225589644e13x^{354}+0.244436067e13x^{356}+0.221143352e13x^{358} \nonumber \\  
&+0.140825874e13x^{360}-0.649324487e11x^{362}-0.221501096e13x^{364}-0.491094324e13x^{366} \nonumber \\  
&-0.784640464e13x^{368}-0.105213812e14x^{370}-0.122555119e14x^{372}-0.122453010e14x^{374} \nonumber \\  
&-0.967388504e13x^{376}-0.387516784e13x^{378}+0.545634834e13x^{380}+0.180323131e14x^{382} \nonumber \\  
&+0.327515035e14x^{384}+0.475474135e14x^{386}+0.593656087e14x^{388}+0.643366501e14x^{390} \nonumber \\  
&+0.582005673e14x^{392}+0.370127270e14x^{394}-0.188373561e13x^{396}-0.586963931e14x^{398} \nonumber \\  
&-0.129986377e15x^{400}-0.207668859e15x^{402}-0.278513129e15x^{404}-0.324482299e15x^{406} \nonumber \\  
&-0.324237662e15x^{408}-0.256042771e15x^{410}-0.102120719e15x^{412}+0.145760298e15x^{414} \nonumber \\  
&+0.480061030e15x^{416}+0.871604193e15x^{418}+0.126546279e16x^{420}+0.158029270e16x^{422} \nonumber \\  
&+0.171287647e16x^{424}+0.154939007e16x^{426}+0.984213300e15x^{428}-0.540858239e14x^{430} \nonumber \\  
&-0.157163820e16x^{432}-0.347711790e16x^{434}-0.555478063e16x^{436}-0.745079732e16x^{438} \nonumber \\  
&-0.868203472e16x^{440}-0.867606804e16x^{442}-0.684883117e16x^{444}-0.272146238e16x^{446} \nonumber \\  
&+0.392956599e16x^{448} \nonumber 
\end{align}

\end{document}